\documentclass[hidelinks]{article} 

\usepackage{authblk}
\usepackage{hyperref}
\usepackage{natbib}
\usepackage[ruled]{algorithm2e}
\usepackage{multirow}
\usepackage{float}
\usepackage{mathtools}
\usepackage{booktabs}
\usepackage{makecell}
\usepackage{siunitx}
\usepackage{xcolor}
\usepackage{ulem} 
\usepackage{comment}
\usepackage{amsthm}
\usepackage[margin=1.5in]{geometry}

\newtheorem{defi}{Definition}[section]
\newtheorem{thm}[defi]{Theorem}

\begin{document}


\title{A Latent Class Modeling Approach for Generating Synthetic Data and Making Posterior Inferences from Differentially Private Counts}

\author[$\dagger$]{Michelle Pistner Nixon}	

\author[$\star$]{Andr\'es F. Barrientos}	

\author[$\ddag$]{Jerome P. Reiter}	

\author[$\mathsection$]{Aleksandra Slavkovi\'{c}}	
\affil[$\dagger$]{\footnotesize College of Information Sciences and Technology, The Pennsylvania State University, University Park, PA }
\affil[$\star$]{\footnotesize Department of Statistics, Florida State University, Tallahassee, FL}
\affil[$\ddag$]{\footnotesize Department of Statistical Science, Duke University, Durham, NC }
\affil[$\mathsection$]{\footnotesize Department of Statistics, The Pennsylvania State University, University Park, PA }




\maketitle

\begin{abstract}
  \noindent 
Several algorithms exist for creating differentially private counts from contingency tables, such as two-way or three-way marginal counts. The resulting noisy counts generally do not correspond to a coherent contingency table, so that some post-processing step is needed if one wants the released counts to correspond to a coherent contingency table.  We present a latent class modeling approach for post-processing differentially private marginal counts that can be used (i) to create differentially private synthetic data from the set of marginal counts, and (ii) to enable posterior inferences about the confidential counts.  We illustrate the approach using a subset of the 2016 American Community Survey Public Use Microdata Sets and the 2004 National Long Term Care Survey. 
\end{abstract}

\section{Introduction}
\label{sec:intro}

National statistical organizations and other data curators, henceforth all called agencies, often seek to share collected information with outside researchers. Doing so requires methods that provide privacy protection to data subjects while maintaining statistical relationships in the data. Many agencies use statistical disclosure control (SDC) methods that blur the private data in a prescribed way, with the aim to provide some level of privacy protection. Traditional SDC methods include  data swapping, cell suppression, top- or bottom- coding, and the addition of random noise \citep{hundepool}. 
Generally, traditional SDC techniques have been applied with low intensity, so as not to degrade the quality of the information in the data.  However, with the growth in available data and powerful computing, many agencies have become concerned that such SDC methods are not adequately protective. 

To address this issue, several agencies have turned to using synthetic data methods \citep{rubin1993discussion,little1993statistical,reiter2005estimating}.  The agency releases data that are generated from some statistical model, estimated with the private data. With a full synthesis, agencies can ensure that no sensitive values are released on the file, which can reduce disclosure risks.  Further, the agency can offer researchers access to record-level data instead of only summary statistics or other high-level information.  Many different methods have been proposed to generate synthetic data \citep[e.g., ][]{reiter2005using, woodcock2009distribution, mateo2004fast,slavkovic2010synthetic} and assess the utility and disclosure risk associated with the data release \citep{drechsler2011synthetic, hundepool,karr2006framework, reiter2014bayesian, snoke}.   
The U.S. Census Bureau has released synthetic data for the Longitudinal Business Database  \citep{synlbd}, Survey of Income and Program Participation  \citep{SIPP}, and the LEHD Origin-Destination Employment Statistics (with the data product being known as \textit{OnTheMap}) \citep{Ashwin2008}. 
Affiliated researchers have proposed methodologies for other national surveys (e.g., federal administrative data from the Office of Personnel Management \citep{barrientos2018providing}), and other statistical agencies have developed synthetic data products such as the Scottish Longitudinal Survey \citep{iCEM} and the German IAB Establishment Data \citep{drechsler2009synthetic}.

Most SDC methods, including many synthetic data techniques, do not provide formally quantifiable privacy guarantees. Instead, agencies evaluate  disclosure risks based on assumptions of intruder knowledge and behavior \citep{hundepool}. An alternative approach is to design and use SDC methods that satisfy differential privacy \citep{dwork2006calibrating}. 
Differentially private algorithms exist for releasing counts \citep[e.g.,] []{dwork2006calibrating, ghosh2012universally}, $k$-way marginals \citep{barak2007privacy, yang2012differential, li2018privacy}, regression coefficients \citep{zhang2012functional,snoke2018pmse, awanslav2021}, and many other quantities that can be viewed as answers to user-specified queries. Several statistical agencies are interested in combining the privacy guarantees from differential privacy with the flexibility afforded by synthetic data. For example, the U.S. Census Bureau uses differentially-private synthetic data as the backbone of its 2020 population census data releases \citep{abowd2018us}. However, it can be challenging to generate differentially private synthetic data with low error  \citep{garfinkel2018issues}.  

In this article, we propose methodology to generate differentially private synthetic data sets for multivariate count data, i.e., data from contingency tables. Our approach can be viewed as a post-processing method for generating synthetic data, a strategy also suggested in  \cite{McKenna_Miklau_Sheldon_2021}. Specifically, we first assume that an agency has selected  a set of marginal counts and has used some existing differentially private algorithm to add noise to those counts.  For example, the agency could select the margins for which accurate values are especially important.  
Alternatively, the agency could select a set of margins that covers many types of analyses done by users of the data, e.g., all $k$-way margins. Second, we specify a Bayesian latent class model for the underlying confidential data \citep{dunson2009nonparametric}. We write the agency-selected margins as functions of the model parameters.  Third, we estimate the parameters in the functions using only the differentially private counts  and a composite likelihood-based approach \citep{lindsay1988composite}. Finally,  we sample from the estimated model to generate record-level, synthetic data. Agencies can generate multiple copies of the synthetic datasets, without any extra privacy loss, in order to enable secondary data analysts to estimate uncertainty and make inferences \citep{reiter:2003}.  Generating differentially-private synthetic data from a user-specified list of summaries can be particularly advantageous for contingency tables with complex structures, e.g., data with structural zeros \citep{manrique2014bayesian, li2018privacy} or nested data such as individuals in households \citep{hu2018dirichlet}.  We illustrate the latter in the online supplementary material.  

The latent class modeling approach also can be viewed as an engine for approximate posterior inferences for the underlying confidential counts given a collection of differentially private counts, under the assumptions implied by the latent class model.  Specifically, analysts can use parameters draws from the Markov chain Monte Carlo (MCMC) sampler to obtain posterior inferences for any functional of the parameters, i.e., any count in the table.

The remainder of the paper is organized as follows. In Section \ref{sec:previous} we review related literature.  In  Section~\ref{sec:prelims} we present preliminary information on data privacy and latent class models. In Section~\ref{sec:propoosed} we discuss the proposed approach and implementation details. In Section~\ref{sec:results},  
we illustrate the approach by making posterior inferences from differentially private counts with a small set of variables from the 2016 American Community Survey (ACS) Public Use Microdata Sets (PUMS), and compare it to several other approaches from the literature.
In Section \ref{sec:NLTCS}, we illustrate the approach for both posterior inference and synthetic data generation using the 2004 National Long Term Care Survey, which has larger dimensions than the ACS example. For both illustrations, we use the Geometric Mechanism \citep{ghosh2012universally} to ensure differential privacy in the first stage counts.  
Finally, in Section~\ref{sec:conclusions} we end with a discussion.

\subsection{Related Literature}\label{sec:previous}

Protection of counts and contingency tables has been the subject of many papers in the privacy literature. Some of the first examples of differentially-private (DP) data were counts and histograms perturbed by the Laplace mechanism~\citep{dwork2006calibrating}. \cite{barak2007privacy} proposed methodology to produce DP marginal counts by using Fourier bases. \cite{yang2012differential} extended this work to multi-dimensional contingency tables and noted some theoretical and practical shortcomings to both approaches. \cite{Ashwin2008} proposed constructing DP contingency tables via a Dirichlet-Multinomial synthesizer, and the Hardt-Ligett-McSherry algorithm \citep{hardt2012simple} used a multiplicative weights approach. 

\cite{park2014pegs} proposed methodology for synthesis of DP contingency tables by  disintegrating the data into suitable building blocks, injecting noise to these blocks, and using a Gibbs sampler to draw synthetic samples. \cite{li2018privacy}  proposed to generate synthetic data by constructing an empirical hashed conditional distribution from the whole histogram and applying a Stability Based Algorithm to these empirical distributions. \cite{mckenna2019graphical} proposed constructing DP synthetic data where ``suitable building blocks'' are a set of selected marginal counts; they add Laplace noise to the selected counts, fit a graphical model to the perturbed counts, and sample synthetic data from this model. Similarly, PrivBayes \citep{zhang2017privbayes} constructs DP synthetic data sets via Bayesian networks. These authors use networks to approximate the underlying data distribution through lower order marginals, add noise to these marginals, and approximate a full data distribution through the noisy counts and constructed Bayesian network.  We note that these two methods placed first \citep{mckenna2021winning} and third \citep{bao2021synthetic}, respectively, in the 2018-2019 National Institute of Standards and Technology Public Safety Communications Research division's \textit{Differential Privacy Synthetic Data Challenge} \citep{nist2018}; see \cite{bowen2019comparative} for a discussion of the performance of these methods in the context of the challenge. Similarly, the CIPHER method  \citep{eugenio2018construction}  estimates the joint distribution of a table based on a lower-order set of DP marginals. Their approach can be viewed as a post-processing technique for generating synthetic data.

\cite{hay2010boosting} showed that accuracy can be greatly improved by accounting for the constraints that a contingency table must satisfy in post-processing steps, although some very recent papers argue for post-processing to be part of posterior modeling (e.g., see \cite{seeman2020private}, and references therin). \cite{lee2015maximum} extended this result by also accounting for the noise distribution and developed a fast generic approach for solving the resulting optimization problem.

\cite{bowen2016comparative} compared many parametric and nonparametric methods for differentially private data synthesis,  beyond those discussed above, and \cite{charest2012empirical} showed that requiring greater privacy can degrade inferential results. \cite{rinott2018confidentiality} discussed several issues related to contingency table release under differential privacy, including the potential effects of rounding and other post-processing steps. They compared several simple mechanisms and offered comments on utility in private data releases. Similarly, \cite{raab2019practical} discussed some practical limitations to contingency table protection under formal privacy guarantees and presented a new method to create differentially private contingency tables from a subset of marginals via the Laplace mechanism  and iterative proportional fitting.

 All of these methods (ours included) are subject to the results of \cite{ullman2020pcps}. They showed that, even in simple cases, it is impossible to construct a polynomial time differentially private algorithm that preserves all two-way marginals.

Accounting for the additional privacy-preserving noise in inference also has been discussed in the literature. \cite{charest2012empirical} used a simple Bayesian model to account for noise added under the Beta-Binomial synthesizer. \cite{karwa2015private} accounted for noise by treating the original data as missing. As the resulting likelihoods are often intractable, they developed a technique relying on a variational approximation for estimation. \cite{seeman2020private} showed that naive post-processing (such as direct modification of counts to meet constraints) can result in loss of information and proposed a Bayesian sampling scheme to post-process counts based on the noisy counts and outside constraining information. \cite{gong2019exact} showed that approximate Bayesian computing can be used to account for DP noise to obtain draws from the correct posterior distribution. \cite{karwa2017sharing} investigated methods to release DP synthetic networks while also accounting for the additional noise by treating the original private data as missing and using a likelihood-based approach. While not dealing with DP tables, \cite{woo2012logistic} proposed several EM algorithms to obtain correct logistic regression estimates for tables that are protected using the Post Randomization Method (PRAM) which could be extended to DP settings.

Turning to latent class models, \cite{si2013nonparametric} used latent class specifications for missing data imputation in contingency tables  of \cite{dunson2009nonparametric}. This latent specification belongs to the family of Bayesian nonparametric techniques, which are well-known for being a flexible modeling choice to capture complex data patterns. \cite{manrique2014bayesian} refined these latent class models to incorporate cases with structural zeros, e.g., a person age five cannot have a college degree. The full table latent class approach was further extended to handle nested variables (e.g., individuals nested in households) \citep{hu2018dirichlet}, structural zeros \citep{akande2019simultaneous}, and missing data \citep{akande2017multiple}; see the Supplementary Material for a  discussion of the model presented in \cite{hu2018dirichlet} and how our proposed approach can be augmented for nested data structures.

\section{Preliminaries}
\label{sec:prelims}

\subsection{Data Privacy Methods}

\subsubsection{Synthetic Data}

Let {$\mathbf{X}=\{(x_{i1}, \dots, x_{ip})\}_{i=1}^n$} be an observed data set comprising $n$ individuals measured on $p$ confidential categorical variables.  For $i=1, \dots, n$ and $j =  1, \dots, p$,  each $x_{ij} \in \{1, \dots, d_j\}$. We refer to the variables in $\mathbf{X}$ using $X_j$ where $j=1,\dots, p$. Synthetic data approaches aim to preserve the joint distribution $f(\mathbf{X})$ of these variables by modeling them simultaneously, e.g., as in \cite{hu2018dirichlet} and \cite{akande2017multiple}, or by using sequential modeling of the form,
 \begin{equation}
 f(\mathbf{X})=f(X_1)f(X_2|X_1) \cdots f(X_p|X_1, \dots, X_{p-1}).
 \label{eq:syn}
 \end{equation} 
 Different types of statistical models can be used for the conditional distributions in (\ref{eq:syn}), including classification and regression trees (CART) \citep{reiter2005using}, random forests \citep{caiola2010random}, kernel density estimators \citep{woodcock2009distribution}, and combinations thereof \citep{barrientos2018providing}. Choice of model depends on both the data structure and desired privacy levels.

\subsubsection{Differential Privacy}

Differential privacy (DP, \citep{dwork2006calibrating, dwork2014algorithmic}) is a formal privacy framework that gaurantees privacy protection by bounding the ratio of output densities for all neighboring data sets. Intuitively, it protects against attackers seeking to learn whether or not any specific individual's data were included (or excluded) in the data set over which the output was calculated.  That is, the inclusion (or exclusion) of a single person's information does not change the output of the algorithm greatly. In this paper we focus on its strongest form, $\epsilon$-DP, where $\epsilon$ is a privacy-loss parameter, with smaller values indicating stronger protection, but our model could be modified for other forms of DP. 

\begin{defi}
 \textbf{$\epsilon$-Differential Privacy \citep{dwork2006calibrating}:}  A randomized {algorithm $\mathcal{A}$} {\rm satisfies } $\epsilon$-differential privacy if for all data sets {$\mathbf{X}=\{(x_{i1}, \dots, x_{ip})\}_{i=1}^n$} and $\mathbf{X^\prime}$ differing on at most one row, and $\mathcal{S} \subseteq \mathrm{Range}(\mathcal{A})$,
 
 \begin{equation}
 \frac{\mathrm{Pr}[\mathcal{A}(\mathbf{X}) \in S]}{\mathrm{Pr}[{\mathcal{A}(\mathbf{X^\prime})} \in S]} \le \exp(\epsilon) \, .
 \end{equation}
\label{def:dp}
\end{defi}

Differential privacy is a property of the algorithm itself. It is achieved by randomness. Many different algorithms have been proposed to release differentially private statistics or data sets (e.g., the Laplace \citep{dwork2006calibrating} or Geometric \citep{ghosh2012universally} mechanisms, which are the most relevant to our seeting). Several relaxations and extensions of differential privacy have been proposed such as $(\epsilon,\delta)$-differential privacy \citep{dwork2006our} and R\'{e}nyi differential privacy \citep{mironov2017renyi}; for more, see \cite{dwork2014algorithmic}.

Many DP mechanisms (ours included) rely on the useful properties of post-processing and sequential composition.  These properties outline how DP privacy guarantee can be preserved and how the privacy-loss parameter is propagated through multiple data releases. Theorem \ref{th:pp_dp} shows that DP is preserved through post-processing, and is key to our proposed approach: the augmented Bayesian latent class model and measurement error model is a post-processing technique for a noisy, privacy-preserving count. Theorem \ref{th:dp_sc} shows how the privacy-loss parameter $\epsilon$ propagates through multiple data releases on the same individual. 

\begin{thm}
Post-processing \citep{dwork2006calibrating}: Let {$\mathcal{A}$} be any randomized algorithm such that {$\mathcal{A}(\mathbf{X})$} is $\epsilon$-differentially private, and let $g$ be any function. Then, {$g(\mathcal{A}(\mathbf{X}))$} also satisfies $\epsilon$-differential privacy.
\label{th:pp_dp}
\end{thm}

\begin{thm}
Sequential composition \citep{mcsherry2009privacy}: Let {$\mathcal{A}_i$} each provide $\epsilon_i$-differential privacy. The sequence of {$\mathcal{A}_i(\mathbf{X})$} provides $(\sum_i \epsilon_i)$-differential privacy.
\label{th:dp_sc}
\end{thm}

\subsubsection{Geometric Mechanism}

The Geometric Mechanism~\citep{ghosh2012universally} adds noise from a two-sided geometric distribution to achieve differential privacy. Properties of this distribution, including simulation techniques, are discussed in \cite{inusah2006discrete} and its references.

\begin{defi} A random variable $Y$ distributed as a two-sided geometric distribution has probability mass function
\begin{equation}
P(Y = k) = \frac{1-\alpha}{1+\alpha} \alpha^{|k|}
\end{equation}
where $0 \le \alpha \le 1$.
\end{defi}

\begin{thm}
Geometric Mechanism \citep{ghosh2012universally}: For $f: \mathcal{D} \rightarrow \mathbf{R}^d$, the mechanism {$\mathcal{A}$} that adds independently drawn noise from a $\mathrm{two}$-$\mathrm{sided}$-$\mathrm{Geom}(\exp\{ \frac{-\epsilon}{\Delta f}\})$ distribution to each of the $d$ terms of satisfies $\epsilon$-differential privacy.
\end{thm}

This mechanism has several appealing properties for protecting count data. First, the added noise values are integers, which eliminates the need for a  post-processing step to deal with decimals. Second, \cite{ghosh2012universally} show that this mechanism is optimal for every potential user regardless of the side information that they posses when releasing DP count queries using a Bayesian framework.

\subsection{Bayesian Latent Class Models}

\cite{dunson2009nonparametric} developed a flexible methodology for modeling unordered categorical data using a Bayesian nonparametric mixture model. Their methods are directly applicable to modeling complete contingency tables.

Suppose each observation $i = 1, \dots, n$ belongs to a latent class denoted by $z_i \in \{1,2, \dots\}$. The probability for each unique combination of variable levels is specified according to these latent classes.  For ease of notation, we allow $x_{ij}$ to stand for random variables as well as the observed data.  The latent class model is of the form,

\begin{equation}
\label{eq:dunson2}
	\begin{split}
		x_{ij} | z_i, \{\Psi_h^{(j)} \}_{h=1}^\infty& \stackrel{ind}{\sim} Multinomial \{1, \Psi_{z_i1}^{(j)} , \dots, \Psi_{z_id_j}^{(j)} \}, i=1, \dots, n, \, j=1, \dots, p,\\
		z_i | \{\pi_h \}_{h=1}^\infty& \stackrel{ind}{\sim} Discrete \{(1, \dots, \infty), (\pi_1, \dots, \pi_\infty)\},\\
		\pi_h &= V_h \prod_{l < h} (1-V_l), \quad   V_h \sim \beta(1,\alpha),\\
		\Psi_h^{(j)} & \sim Dirichlet(a_{j1}, \dots, a_{{jd}_{j}}),\\
	\end{split}
\end{equation}
where $\Psi_h^{(j)} = ( \Psi_{h1}^{(j)}, \dots , \Psi_{hd_j}^{(j)} )$, $\alpha > 0$, and $(a_{j1}, \dots, a_{jdj})$ is a vector with positive components. Under model (\ref{eq:dunson2}), for any feasible set of values $(c_1, \dots, c_p)$ of the categorical variables, one has
\begin{equation}
Pr(x_{i1} = c_1, \dots, x_{ip}=c_p|\{\Psi_h^{(j)}\}_{h=1}^\infty, \{\pi_h\}_{h=1}^\infty) = \sum_{h=1}^\infty \pi_h \prod_{j=1}^p \Psi_{hc_j}^{(j)}, \,
\label{eq:probLCM}
\end{equation}
which corresponds to a mixture of product multinomials with a Dirichlet process \citep{ferguson1973bayesian} as a mixing distribution.

To facilitate posterior computation, we use a finite dimensional approximation of model (\ref{eq:dunson2}), which can be estimated using Gibbs sampling. The approximation relies on the assumption that $z_i \in \{1, \dots, k\}$, where $k < \infty$, which is equivalent to truncating the mixture model (\ref{eq:probLCM}) to $k$ components; see \cite{ishwaran2001gibbs} for a discussion of the truncated Dirichlet processes. Similar approximations along with Gibbs sampling have been successfully used for private data synthesis, including in \cite{hu2018dirichlet}, \cite{manrique2014bayesian} and \cite{akande2017multiple}, among others.

\subsection{Composite Likelihood Methods}

Composite likelihood methods allow analysts to  circumvent the specification of the full joint likelihood function \citep{lindsay1988composite,varin2011overview}. Instead, analysts can use an approximation to the likelihood function, often with the assumption of some form of independence, to simplify computation. 

We use composite likelihood methods to approximate the joint likelihood of the latent class model given an agency-specified set of marginal counts.  Let $(M_1, \ldots, M_T)$ represent the list of marginal counts of $\mathbf{X}$ with $\theta = \left( \{\Psi_h^{(j)} \}_{h=1,j=1}^{k,p}, \, \{\pi_h \}_{h=1}^k\right)$. Let $\mathcal{L}_t(\theta; M_t)$ be the likelihood function corresponding to $M_t$ induced by Model \eqref{eq:dunson2}. We approximate the joint likelihood of $\theta$ given $(M_1, \ldots, M_T)$ using 
\begin{equation}
    \mathcal{L}_C(\theta;M_1, \ldots, M_T) \approx \prod_{t=1}^T \mathcal{L}_t(\theta; M_t). 
\end{equation}
The composite likelihood $\mathcal{L}_C$ allows for inference about $\theta$ only from the marginal counts \citep{varin2011overview}. We note that, while computationally expedient, inference based on only a subset of counts of $\mathbf{X}$ could result in inaccurate estimates of $\theta$ \citep{walker2013bayesian}, particularly for regions of its distribution that describe the counts excluded from the set of selected marginals. Furthermore, not all combinations of marginals, and possibly conditional distributions fully and uniquely specify the joint distribution (e.g., see \cite{fienberg2005preserving, slavzhupet2015} within the context of confidentiality protection, and more general related references therein). 

\section{Proposed Methods}
\label{sec:propoosed}

\subsection{Overview}

We assume the agency adds noise to the true counts for a set of agency-specified marginals, $(M_1, \ldots, M_T)$, computed from $\mathbf{X}$ using the Geometric mechanism.   In our simulation examples, we use the definition of sensitivity based on changing one row; hence, the sensitivity for each $M_t$, where $t=1, \ldots, T$, equals two.\footnote{Under other definitions of sensitivity, the sensitivity equals 1 since adding or deleting one row changes at most one marginal count.} We assume the agency has  determined appropriate values of $\epsilon$ to control the overall privacy budget. See Section~\ref{sec:results} for a discussion in the context of our simulations.  

When estimating the parameters of the Bayesian latent class model, we account for the additional noise due to DP using a measurement error model \citep{fuller2009measurement}. Specifically, we treat the true underlying counts as unknown and incorporate the privacy preserving mechanism into the model. Let $\mathbf{\Tilde{M}}$ denote the observed noisy marginal counts. We have 
\begin{equation}
\begin{split}
\mathbf{\Tilde{M}} & = (M_1+\varepsilon_1,\ldots,M_T+\varepsilon_T),\\ \varepsilon_t & \overset{ind}{\sim} 
\mathrm{two}\mbox{-}\mathrm{sided}\mbox{-}\mathrm{Geom}_{r_t}\left(\exp\left\{ \frac{-\epsilon}{\Delta M_t T}\right\}\right), t=1,\ldots,T,
\\
M_t |\boldsymbol{\pi}_k, \boldsymbol{\Psi}_k   & \overset{ind}{\sim} {\rm Multinomial}_{r_t}(n,P_t(\boldsymbol{\pi}_k, \boldsymbol{\Psi}_k)), \;  t=1,\ldots,T,  \\ \boldsymbol{\pi}_k&=\{\pi_h\}_{h=1}^k, \; \boldsymbol{\Psi}_k = \{\Psi_h^{(j)}\}_{h=1,j=1}^{k,p} , 
\end{split}
\label{eq:model}
\end{equation}
where each $M_t$ comprises $r_t$ counts, $\varepsilon_t$ is a random vector with $r_t$ independent components distributed as $\mathrm{two}\mbox{-}\mathrm{sided}\mbox{-}\mathrm{Geom}\left(\exp\left\{ \frac{-\epsilon}{\Delta M_t}\right\}\right)$, and ${\Delta M_t}=2$ is the sensitivity of $M_t$. We complete the model specification by assuming the prior distributions for $\boldsymbol{\pi}_k$ and $\boldsymbol{\Psi}_k$ used in (\ref{eq:dunson2}). Since the summaries $M_t$ are assumed to be marginal counts, the distribution induced by $\mathbf{X} \mapsto M_t$ is multinomial with parameters $n$ and $P_t(\boldsymbol{\pi}_k, \boldsymbol{\Psi}_k)$, where $P_t(\boldsymbol{\pi}_k, \boldsymbol{\Psi}_k)$ represents the probabilities of the counts in $M_t$ and is computed using expression (\ref{eq:probLCM}).

The third line of \eqref{eq:model} represents our ``augmented'' step \citep{lindsay1988composite,varin2011overview}: each marginal count is assumed to be independently drawn from multinomial distributions.  We rely on this assumption as it is not obvious how to characterize the joint distribution for $(M_1, \ldots, M_T) |\boldsymbol{\pi}_k, \boldsymbol{\Psi}_k$. Although the premise of independence leads to a product of multinomial distributions that places positive probability outside the range of $(M_1, \ldots, M_T)$, we expect such probability to be small since $(P_1(\boldsymbol{\pi}_k,  \boldsymbol{\Psi}_k), \ldots, P_T(\boldsymbol{\pi}_k,  \boldsymbol{\Psi}_k))$ is a coherent system of probabilities that accounts for the underlying constraints among  $M_1, \ldots, M_T$. Hence, in order to generate synthetic tables that consider such constraints, our strategy is to sample from the distribution of $\boldsymbol{\pi}_k,  \boldsymbol{\Psi}_k|\boldsymbol{\Tilde{M}}$, and approximate the underlying table cell probabilities by
\begin{eqnarray*}
& & \hspace{-12mm} Pr(x_{{(n+1)}1} = c_1, \dots, x_{{(n+1)}p}=c_p|\boldsymbol{\Tilde{M}}) = \\ 
& & \int\int  Pr(x_{{(n+1)}1} = c_1, \dots, x_{{(n+1)}p}=c_p|\boldsymbol{\pi}_k, \boldsymbol{\Psi}_k)
Pr(\boldsymbol{\pi}_k, \boldsymbol{\Psi}_k|\boldsymbol{\Tilde{M}})d\boldsymbol{\pi}_k d\boldsymbol{\Psi}_k.
\end{eqnarray*}
Here, $Pr(x_{{(n+1)}1} = c_1, \dots, x_{{(n+1)}p}=c_p|\boldsymbol{\pi}_k, \boldsymbol{\Psi}_k)$ is defined as in (\ref{eq:probLCM}). 

Regardless of which marginal counts are used in estimation, probabilities corresponding to any cell of the contingency table or marginal count can be estimated through ($\boldsymbol{\pi}_k, \boldsymbol{\Psi}_k$). At each iterate, the desired probabilities can be tabulated using the sampled values of ($\boldsymbol{\pi}_k, \boldsymbol{\Psi}_k$) at that iterate. This results in a posterior distribution for each desired probability. Synthetic data can be created from the full cell counts either by straightforward post-processing (e.g., multiply full table probabilities by the desired table size and expand)  or by multinomial sampling of the  cells weighted by the full table probabilities.

Any arbitrary set of summary statistics can be fed into the model. However, to increase the statistical usefulness of the estimated table, we suggest using sets of summary statistics that include at least one instance of every variable and its corresponding levels, which we do here. In addition, when privacy budgets are constrained, we recommend using few rather than many marginal counts. Using a smaller set of marginals allocates more of the privacy budget to each marginal table, which can improve accuracy for those counts. We note that  
the dimension of parameters ($\boldsymbol{\pi}, \boldsymbol{\psi}$) to be estimated does not depend on the number of tables passed to the model.

In general, we expect these results to improve as the accuracy of the formally private counts improves. However, difficulties can arise with the measurement error step of our model if it is hard to specify the DP noise generating process.

This approach to differentially private synthesis does not suffer from privacy loss from running the estimation algorithm.  Noise is injected to the marginal counts prior to the MCMC sampling \citep{chaudhuri2013near}. The algorithm uses these noisy marginals and knowledge of the noise distribution (which does not violate differential privacy) to estimate the underlying contingency table through ($\boldsymbol{\pi}_k, \boldsymbol{\Psi}_k)$.

\subsection{Illustrative Implementation}

To illustrate implementations, suppose the agency has generated all 40 two-way marginal counts from  a $2^5$ table. We can write each two-way marginal probability as a sum over all combinations of the remaining three variables, for example, 
\begin{equation}
\begin{split}
    Pr(x_{i1} = 0,x_{i2}=0|\boldsymbol{\pi}_k, \boldsymbol{\Psi}_k) &= \sum_{j=0}^1 \sum_{l=0}^1 \sum_{m=0}^1 Pr(x_{i1} = 0,x_{i2} = 0, x_{i3} = j, x_{i4} = l, x_{i5} = m|\boldsymbol{\pi}_k, \boldsymbol{\Psi}_k) \\
    &= \sum_{h=1}^k \pi_h  \Psi_{h0}^{(1)} \Psi_{h0}^{(2)} \sum_{j=0}^1 \sum_{l=0}^1 \sum_{m=0}^1  \Psi_{hj}^{(3)} \Psi_{hl}^{(4)} \Psi_{hm}^{(5)}\, .
\end{split}
\label{eq:probLCM3}
\end{equation}
Similarly, we have  
\begin{equation}
\begin{split}
    Pr(x_{i1} = 0,x_{i2}=1|\boldsymbol{\pi}_k, \boldsymbol{\Psi}_k) &= 
     \sum_{h=1}^k \pi_h  \Psi_{h0}^{(1)} \Psi_{h1}^{(2)} \sum_{j=0}^1 \sum_{l=0}^1 \sum_{m=0}^1  \Psi_{hj}^{(3)} \Psi_{hl}^{(4)} \Psi_{hm}^{(5)}\\
    Pr(x_{i1} = 1,x_{i2}=0|\boldsymbol{\pi}_k, \boldsymbol{\Psi}_k) &= 
     \sum_{h=1}^k \pi_h  \Psi_{h1}^{(1)} \Psi_{h0}^{(2)} \sum_{j=0}^1 \sum_{l=0}^1 \sum_{m=0}^1  \Psi_{hj}^{(3)} \Psi_{hl}^{(4)} \Psi_{hm}^{(5)}\\
         Pr(x_{i1} = 1,x_{i2}=1|\boldsymbol{\pi}_k, \boldsymbol{\Psi}_k) &= 
     \sum_{h=1}^k \pi_h  \Psi_{h1}^{(1)} \Psi_{h1}^{(2)} \sum_{j=0}^1 \sum_{l=0}^1 \sum_{m=0}^1  \Psi_{hj}^{(3)} \Psi_{hl}^{(4)} \Psi_{hm}^{(5)}.\\
\end{split}
\label{eq:probLCM4}
\end{equation}
We denote the probabilities in \eqref{eq:probLCM3} and \eqref{eq:probLCM4} by $P_t(\boldsymbol{\pi}_k, \boldsymbol{\Psi}_k)$ for each corresponding count in Model \ref{eq:model}.   

After writing all the two-way marginal counts as functions of the latent class parameters, we can estimate the posterior distribution of all parameters using a MCMC sampler; see Supplementary Section 2 for details. We developed an \texttt{R} package to fit {the proposed model} given an arbitrary set of two-way marginals from a $2^p$ contingency table of any size\footnote{This package, along with all code used in this paper can be found on Github at \url{https://github.com/michellepistner/BayesLCM}..

One source of computational overhead for the MCMC sampler is computing all the marginal probabilities from the model parameters, i.e., $P_t(\boldsymbol{\pi}_k, \boldsymbol{\Psi}_k)$. The complexity of this calculation depends on the size of the contingency table, the number of latent classes, and the total number of marginals used as input. In simulations with our \texttt{R} package, we find that computing each two-way marginal probability is extremely fast ($<2$ milliseconds) when the dimension is ten or less regardless of the number of latent classes. However, run time greatly increases for higher dimensions, leading to slower sampling for the overall model for higher-dimensional tables. Full results are presented in Section 3 of the supplementary material.}

\section{Application to American Community Survey Data}
\label{sec:results}

In this section, we illustrate the proposed approach using data from the ACS PUMS and a small number of variables; see Table~\ref{tab:rules}. In this illustration, we use the latent class model as a post-processing algorithm to obtain posterior inferences from a set of released noisy counts.

\subsection{Data Description}
\label{sec:acs}

For each simulation run, we randomly select a subset of 10,000 individuals from the 2016 one-year ACS PUMS. We use five variables on each individual, namely sex, age, race, citizenship, and income. We recode the latter four variables to binary variables using the rules in Table~\ref{tab:rules}.  Table \ref{tab:2wayMargins} displays all the 2-way marginal tables from one such sample, to give a sense of the typical counts. 

\begin{table}[ht]
    \caption{Data description for the ACS recoded variables. Original data was collected from the 2016 1-year ACS PUMS and recoded according to the guidelines below.}
    \centering 
        \begin{tabular}{l l  l}
            \textbf{Variable} & \textbf{Label} & \textbf{Levels} \\
            \hline
            Sex & SEX & 0: males, 1: females\\ 
            Age & AGE&  0: under age 18, 1: over age 18\\
            Race &RACE& 0: non-white, 1: white\\
            Citizenship &CIT& 0: non-U.S. citizen, 1: U.S. citizen\\
            Income&INC& 0: income under federal poverty line\\
            && 1: income above federal poverty line\\
            \hline
        \end{tabular}
    \label{tab:rules}
\end{table}

\begin{table}[ht]
    \centering
    \caption{All two-way marginal tables for one randomly sampled subset of 10,000 observations from the 2016 ACS PUMS.} 
    \footnotesize
    \begin{tabular}{l l l}
         &  \multicolumn{2}{c}{Age}\\
        \cmidrule{2-3}
         Citizenship &  0 & 1\\
         \hline
         0 & 11 & 596\\
         1 & 443 & 8,950\\
         \hline
    \end{tabular}
    \hspace{1in}
        \begin{tabular}{l l l}
         &  \multicolumn{2}{c}{Race}\\
        \cmidrule{2-3}
         Citizenship &  0 & 1\\
         \hline
         0 & 299 & 308\\
         1 & 1,731 & 7,662\\
         \hline
    \end{tabular}
        \bigskip
    
        \begin{tabular}{l l l}
         &  \multicolumn{2}{c}{Sex}\\
        \cmidrule{2-3}
         Citizenship &  0 & 1\\
         \hline
         0 & 273 & 354\\
         1 & 4,505 & 4,888\\
         \hline
    \end{tabular}
    \hspace{1in}
        \begin{tabular}{l l l}
         &  \multicolumn{2}{c}{Income}\\
        \cmidrule{2-3}
         Citizenship &  0 & 1\\
         \hline
         0 & 294 & 313\\
         1 & 2,916 & 6,477\\
         \hline
    \end{tabular}
    
        \bigskip
    
        \begin{tabular}{l l l}
         &  \multicolumn{2}{c}{Race}\\
        \cmidrule{2-3}
         Age &  0 & 1\\
         \hline
         0 & 110 & 344\\
         1 & 1,920 & 7,626\\
         \hline
    \end{tabular}    
    \hspace{1in}
        \begin{tabular}{l l l}
         &  \multicolumn{2}{c}{Sex}\\
        \cmidrule{2-3}
         Age &  0 & 1\\
         \hline
         0 & 239 & 215\\
         1 & 4,539 & 5,007\\
         \hline
    \end{tabular}
        \hspace{1in}
        \begin{tabular}{l l l}
         &  \multicolumn{2}{c}{Income}\\
        \cmidrule{2-3}
         Age &  0 & 1\\
         \hline
         0 & 445 & 9\\
         1 & 2,765 & 6,781\\
         \hline
    \end{tabular}
    
        \bigskip
    
        \begin{tabular}{l l l}
         &  \multicolumn{2}{c}{Sex}\\
        \cmidrule{2-3}
         Race &  0 & 1\\
         \hline
         0 & 945 & 1,085\\
         1 & 3,833 & 4,137\\
         \hline
    \end{tabular}
        \hspace{1in}
    \begin{tabular}{l l l}
         &  \multicolumn{2}{c}{Income}\\
        \cmidrule{2-3}
         Race &  0 & 1\\
         \hline
         0 & 827 & 1,203\\
         1 & 2,382 & 5,587\\
         \hline
    \end{tabular}
    \hspace{1in}
    \begin{tabular}{l l l}
         &  \multicolumn{2}{c}{Income}\\
        \cmidrule{2-3}
         Sex &  0 & 1\\
         \hline
         0 & 1,281 & 3,497\\
         1 & 1,929 & 3,293\\
         \hline
    \end{tabular}
    \label{tab:2wayMargins}
\end{table}

\subsection{Simulation Details}

We base our example on the assumptions that the agency has decided that all two-way marginal tables are useful to the end data user and will create differentially private versions of them.  To make the differentially private counts, we add independent noise drawn from the Geometric mechanism with $\epsilon \in \{0.25, 0.5, 1.0\}$. We invoke sequential composition theorems to adhere to the specified values of $\epsilon$. In particular, each of the 40 two-way marginal counts belongs to one of $5\choose 2$  $= 10$ two-way tables. Thus, each individual contributes to only one cell per table but contributes across all ten tables. Accordingly, each marginal table should be perturbed with one-tenth of the privacy budget.

After generating the noisy counts in each simulation run, we use a MCMC sampler to iteratively sample the unknown parameters, $(M_1, \ldots, M_T,  \boldsymbol{\pi}_k, \boldsymbol{\Psi}_k)$. Unless otherwise noted, we set the number of latent classes equal to $k=10$. In fitting, generally at least four classes have non-trivial mass (as determined by $\boldsymbol{\pi_k}$); see Supplementary Section 2 for model implementation details.

We run the algorithm for 5,000 iterations, discarding the first 2,000 iterations as burn-in. We monitor convergence of model probabilities using trace plots, a standard practice in Bayesian data analysis \citep{gelman2013bayesian, roy2020convergence}.\footnote{Due to {potential identifiability issues due to label-switching \cite{gelman2013bayesian}}, monitoring convergence of $(\boldsymbol{\pi}_k, \boldsymbol{\Psi}_k)$ is infeasible. Instead, we monitor estimated full and marginal probabilities.} In each iteration, we compute the probabilities for all two-way margins and cells in the full table according to the latent class specification using the {sampled parameters} $(\boldsymbol{\pi}_k, \boldsymbol{\Psi}_k)$ {at that iterate}.  

We compare outputs from the latent class model to output generated using PrivBayes \citep{zhang2017privbayes}, following the architecture for adapting PrivBayes to private data releases described in \cite{ping2017datasynthesizer}. To do so, we adapt their publicly-available Python code, accessible at \url{https://github.com/DataResponsibly/DataSynthesizer}. We also compare to a graphical model-based approach \citep{mckenna2019graphical} by adapting their publicly-available Python code, accessible at \url{https://github.com/ryan112358/private-pgm}. We chose these methods due to their prevalence, performance capabilities, and similarities to our proposed approach in terms of inputs in the sense that all comparison methods are created from marginal counts instead of the full, underlying table. 

The codes for these two methods return different outputs. PrivBayes returns a synthetic data set, which we convert into probabilities for the full table.  The graphical model-based approach returns a vector of numbers (not necessarily counts)  summing to the sample size that corresponds to ``counts'' for each of the full table cells.  We normalize these numbers to create  probabilities for the full table. We note that, unlike our proposed latent class approach, neither of these methods generate uncertainty estimates.   

\subsection{Results}

Before turning to the results using differentially private marginals, we first study the usefulness of the composite likelihood approach and the corresponding assumption of independent counts absent privacy concerns, i.e., with $\epsilon=\infty$.   
\subsubsection{Performance Absent Privacy Concerns}

We estimate the augmented latent class model for the underlying table without adding privacy-preserving noise to the marginal counts. To do, we fit model \eqref{eq:model} using the true $(M_1, \dots, M_T)$ and without the measurement error components in the first two lines of that model.

\begin{figure}[htbp]
	\centering
		\includegraphics[width=3.5in]{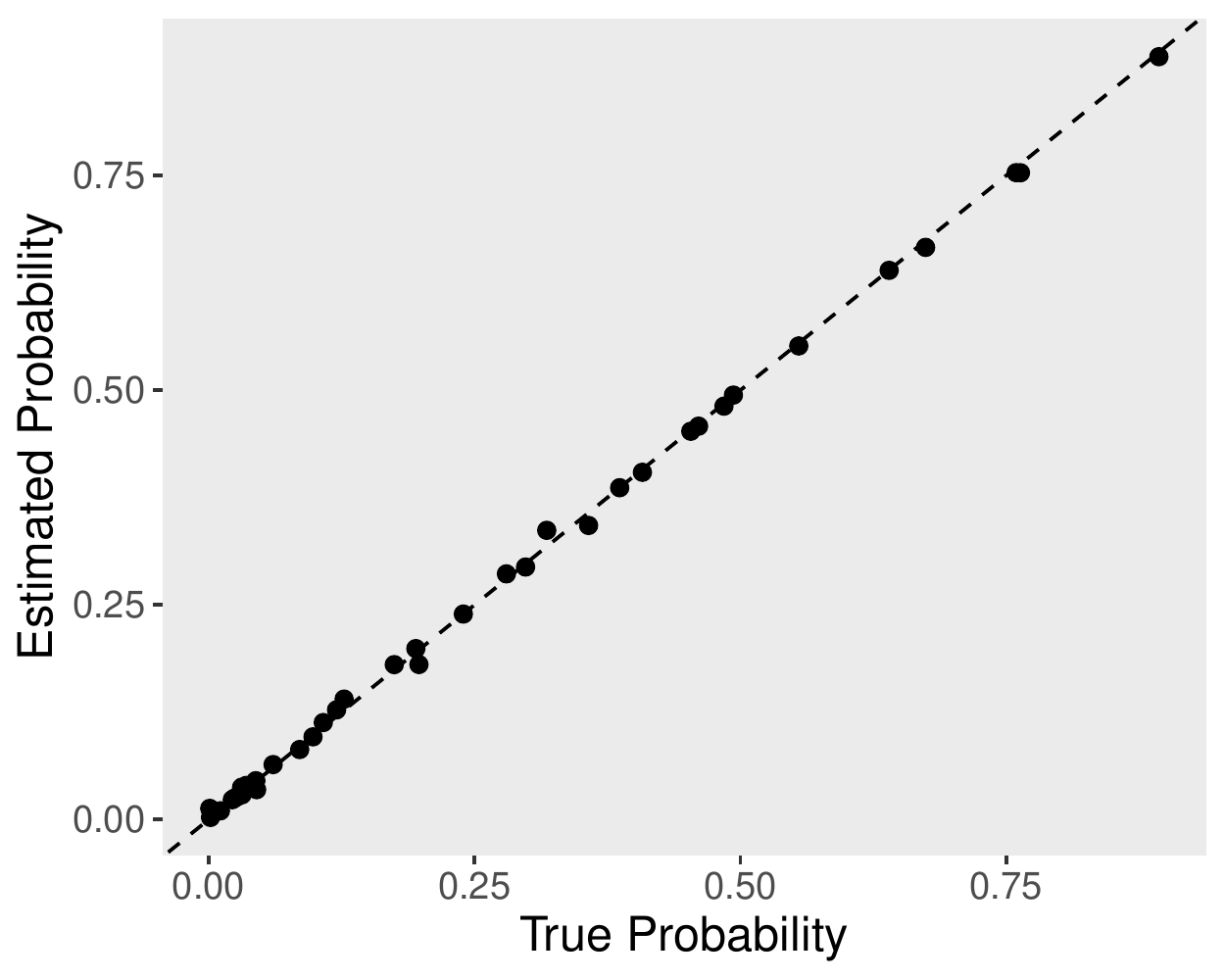}\\
	\includegraphics[width=3.5in]{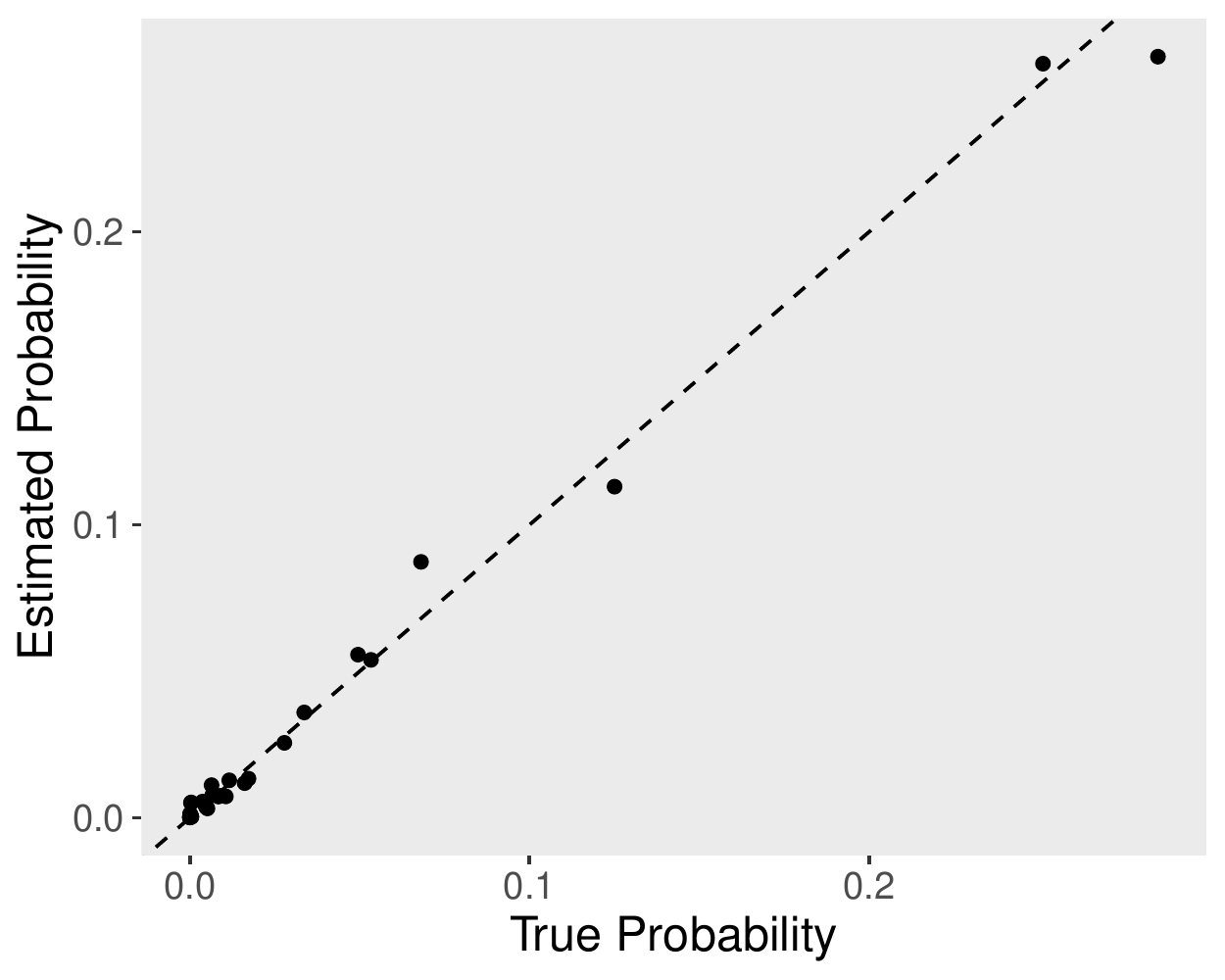}
	\caption{True versus estimated {two-way} marginal (top) and full table {(bottom)} probabilities for the ACS simulation with no noise added for privacy. True counts without added noise used as input.  The number of latent classes  set to 10. As two-way margins were used as inputs, we see more accurate estimation of the marginal probabilities. Accurate estimation of the full cell probabilities is not guaranteed, but can occur.}
	\label{fig:noPrivacy}
\end{figure}

Figure~\ref{fig:noPrivacy} displays posterior modes from one run of the simulation for the marginal and full table probabilities.  Results from additional runs are very similar. Both marginal and full cell probabilities are estimated accurately. The independence assumption appears to have minimal negative effects on estimation of the marginal probabilities, while also returning reasonable estimates for the full table cell probabilities.  We note that this latter fact stems from the absence of strong three-way (and higher) interaction effects among the variables.  In general, one  should not expect accurate estimates for the full table; rather, one should expect accurate estimates of the marginal counts used in the estimation routines.

\subsubsection{Performance with Noisy Counts}

We next generate differentially-private counts and corresponding posterior inferences for the data described in Section~\ref{sec:acs}, using the algorithm described in Section~\ref{sec:propoosed}.

\begin{figure}
	\centering
	\includegraphics[width=3.5in]{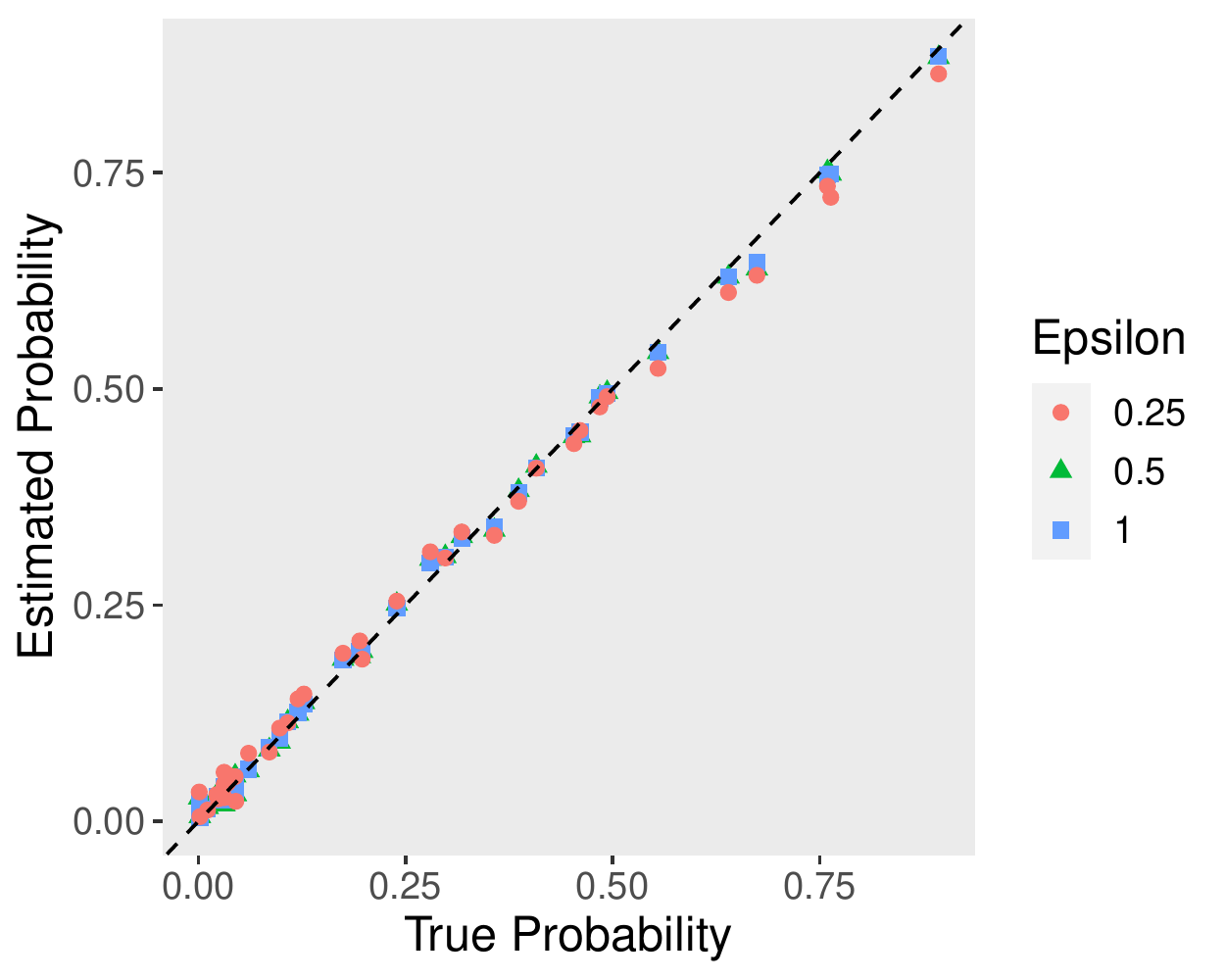}\\
	\includegraphics[width=3.5in]{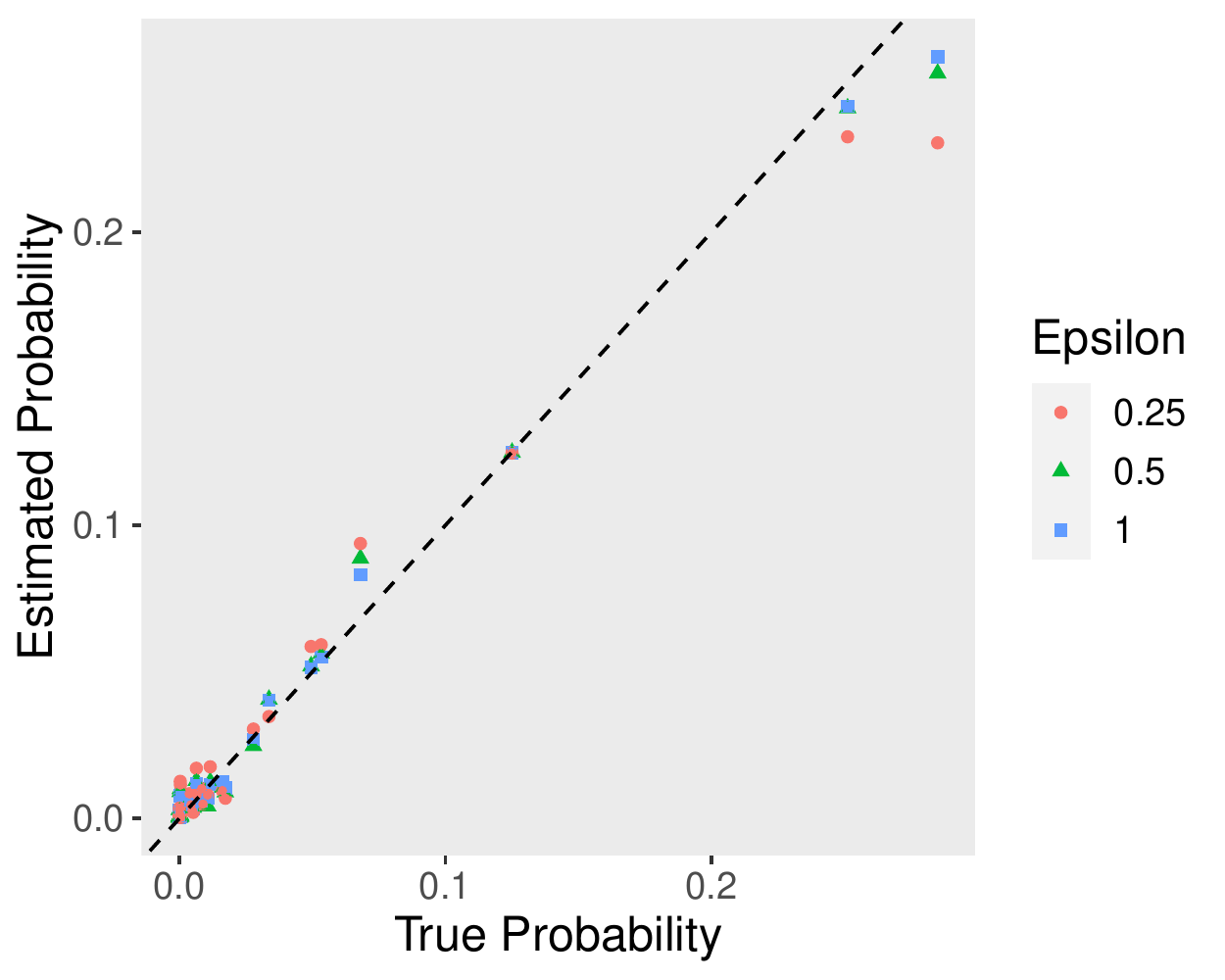}
	\caption{Plots of true marginal probabilities versus estimated marginal probabilities (top) and true full table probabilities versus estimated probabilities (bottom) for $\epsilon \in \{0.25, 0.5, 1.0$\}. Estimated marginal probabilities denote the mean of the posterior distribution for the corresponding marginal probability. As expected, estimation accuracy visually increase as the $\epsilon$ grows.}
	\label{fig:graphs1}
\end{figure}

Figure~\ref{fig:graphs1} displays posterior means for the estimated marginal and full cell probabilities
modes from one run of the simulation.  Results from additional runs are very similar. The results exhibit some expected trends. As the privacy loss budget $\epsilon$ increases, the amount of added noise decreases, resulting in better estimates of marginal and full cell probabilities. In the case of marginal probabilities, results for all values of the privacy loss budget return estimates that coincide well with the underlying probabilities, as shown in Figure~\ref{fig:graphs1}. The results demonstrate this behavior for all values of $\epsilon$. For the full table probabilities, the proposed approach reasonably reconstructs the underlying probabilities even though full table counts are not used in estimation. In general, there is more variation from the truth when compared to the estimates of the marginal probabilities. This also corresponds with the privacy budget: lower values of $\epsilon$ display more disagreement from the true values compared to higher values. This is reasonable to expect as only the marginal counts were used directly in estimation.

We also conducted repeated simulations to assess how well the procedure captures the underlying marginal counts for each value of $\epsilon$. To do so, we repeat the entire procedure 100 times.   Each time we sample 10,000 observations from the full original data, compute all two-way marginals, apply the Geometric Mechanism, and implement our method. For each marginal count, we examine the posterior distribution of the corresponding marginal probability and construct 95\% credible intervals  using the 2.5$\mathrm{^{th}}$ and 97.5$\mathrm{^{th}}$ quantiles of the posterior distribution. 
We record the coverage of the true probability, that is, whether or not the population count is inside the posterior interval. 

\begin{table}[!htbp] \centering 
      \caption{Coverage probabilities (``Cov.'')and average credible interval length (``Length'') for the forty two-way marginal counts.}
    \begin{tabular}{l l l l l l l l l l}
             & & \multicolumn{2}{c}{\boldmath{$\epsilon$ = 0.25}} &
        \multicolumn{2}{c}{\boldmath{$\epsilon$ = 0.50}}& \multicolumn{2}{c}{\boldmath{$\epsilon$ = 1.00}} & \multicolumn{2}{c}{\boldmath{$\epsilon$ = $\infty$}}\\
        \cmidrule{3-10}
        Cell & Pop. Value & \textit{Cov.} & \textit{Length} & \textit{Cov.} & \textit{Length} & \textit{Cov.} & \textit{Length} & \textit{Cov.} & \textit{Length}\\
        \hline
        \hline
        00... & $0.002$ & $0.10$ & $0.01$ & $0.44$ & $0.01$ & $0.70$ & $0.01$ & $0.97$ & $0.01$ \\ 
        01... & $0.061$ & $0.97$ & $0.04$ & $1.00$ & $0.03$ & $1.00$ & $0.02$ & $1.00$ & $0.02$ \\ 
        10... & $0.044$ & $0.71$ & $0.03$ & $0.90$ & $0.03$ & $1.00$ & $0.02$ & $1.00$ & $0.03$ \\ 
        11... & $0.894$ & $0.49$ & $0.05$ & $0.89$ & $0.04$ & $0.99$ & $0.03$ & $1.00$ & $0.04$ \\ 
        0.0.. & $0.031$ & $0.83$ & $0.03$ & $0.89$ & $0.03$ & $0.98$ & $0.03$ & $1.00$ & $0.03$ \\ 
        0.1.. & $0.031$ & $0.16$ & $0.04$ & $0.37$ & $0.03$ & $0.74$ & $0.02$ & $0.92$ & $0.02$ \\
        1.0.. & $0.174$ & $0.88$ & $0.05$ & $0.89$ & $0.04$ & $0.96$ & $0.04$ & $0.99$ & $0.04$\\ 
        1.1.. & $0.763$ & $0.57$ & $0.06$ & $0.79$ & $0.05$ & $0.95$ & $0.04$ & $0.97$ & $0.04$\\ 
        0..0. & $0.031$ & $1.00$ & $0.03$ & $1.00$ & $0.03$ & $1.00$ & $0.02$ & $1.00$ & $0.02$ \\ 
        0..1. & $0.031$ & $0.98$ & $0.03$ & $1.00$ & $0.03$ & $1.00$ & $0.02$ & $1.00$ & $0.02$\\ 
        1..0. & $0.453$ & $1.00$ & $0.05$ & $1.00$ & $0.05$ & $1.00$ & $0.04$ & $1.00$ & $0.04$ \\ 
        1..1. & $0.485$ & $0.99$ & $0.05$ & $1.00$ & $0.05$ & $1.00$ & $0.04$ & $1.00$ & $0.04$\\ 
        0...0 & $0.027$ & $1.00$ & $0.03$ & $1.00$ & $0.03$ & $1.00$ & $0.03$ & $1.00$ & $0.03$ \\ 
        0...1 & $0.035$ & $0.88$ & $0.03$ & $0.98$ & $0.03$ & $1.00$ & $0.02$ & $1.00$ & $0.02$ \\ 
        1...0 &$0.298$ & $1.00$ & $0.06$ & $1.00$ & $0.05$ & $1.00$ & $0.05$ & $1.00$ & $0.06$ \\ 
        1...1 &$0.640$ &$0.98$ & $0.06$ & $1.00$ & $0.06$ & $1.00$ & $0.05$ & $1.00$ & $0.06$\\ 
        .00.. &$0.011$ & $0.94$ & $0.02$ & $0.96$ & $0.02$ & $1.00$ & $0.02$ & $1.00$ & $0.02$ \\ 
        .01.. & $0.035$ & $0.77$ & $0.03$ & $0.95$ & $0.03$ & $1.00$ & $0.02$ & $1.00$ & $0.03$ \\ 
        .10..& $0.195$ & $0.99$ & $0.05$ & $1.00$ & $0.04$ & $1.00$ & $0.04$ & $1.00$ & $0.04$\\ 
        .11.. & $0.759$ & $0.89$ & $0.06$ & $0.97$ & $0.05$ & $0.99$ & $0.04$ & $1.00$ & $0.05$ \\ 
        .0.0. & $0.024$ & $0.95$ & $0.03$ & $1.00$ & $0.02$ & $1.00$ & $0.02$ & $1.00$ & $0.02$ \\ 
        .0.1. & $0.022$ & $0.78$ & $0.03$ & $0.84$ & $0.03$ & $0.93$ & $0.02$ & $1.00$ & $0.02$\\ 
        .1.0. & $0.461$ & $1.00$ & $0.05$ & $1.00$ & $0.05$ & $1.00$ & $0.04$ & $1.00$ & $0.05$ \\ 
        .1.1. & $0.493$ & $0.98$ & $0.05$ & $0.99$ & $0.05$ & $1.00$ & $0.04$ & $1.00$ & $0.05$ \\ 
        .0..0 & $0.045$ & $0.60$ & $0.03$ & $0.67$ & $0.03$ & $0.93$ & $0.03$ & $1.00$ & $0.04$ \\ 
        .0..1 & $0.001$ & $0.00$ & $0.03$ & $0.00$ & $0.02$ & $0.00$ & $0.02$ & $0.00$ & $0.02$ \\ 
        .1..0 & $0.280$ & $0.93$ & $0.06$ & $0.90$ & $0.05$ & $0.93$ & $0.05$ & $1.00$ & $0.06$\\ 
        .1..1 & $0.674$ & $0.21$ & $0.06$ & $0.43$ & $0.06$ & $0.70$ & $0.05$ & $0.93$ & $0.06$ \\ 
        ..00. & $0.098$ & $1.00$ & $0.06$ & $1.00$ & $0.05$ & $1.00$ & $0.04$ & $1.00$ & $0.04$\\ 
        ..01. & $0.108$ & $1.00$ & $0.06$ & $1.00$ & $0.05$ & $1.00$ & $0.04$ & $1.00$ & $0.04$\\ 
        ..10. & $0.386$ & $1.00$ & $0.06$ & $1.00$ & $0.05$ & $1.00$ & $0.05$ & $1.00$ & $0.05$ \\ 
        ..11. & $0.408$ & $1.00$ & $0.06$ & $1.00$ & $0.05$ & $1.00$ & $0.05$ & $1.00$ & $0.05$\\ 
        ..0.0 &$0.086$ & $1.00$ & $0.06$ & $1.00$ & $0.05$ & $1.00$ & $0.05$ & $1.00$ & $0.05$\\ 
        ..0.1 &$0.120$ & $0.99$ & $0.05$ & $1.00$ & $0.05$ & $1.00$ & $0.05$ & $1.00$ & $0.04$  \\ 
        ..1.0 & $0.239$ & $1.00$ & $0.06$ & $1.00$ & $0.06$ & $1.00$ & $0.05$ & $1.00$ & $0.06$\\ 
        ..1.1 & $0.555$ & $1.00$ & $0.07$ & $1.00$ & $0.06$ & $1.00$ & $0.06$ & $1.00$ & $0.06$\\ 
        ...00 & $0.127$ & $1.00$ & $0.06$ & $1.00$ & $0.06$ & $1.00$ & $0.05$ & $1.00$ & $0.05$\\ 
        ...01 & $0.357$ & $1.00$ & $0.07$ & $1.00$ & $0.06$ & $1.00$ & $0.05$ & $1.00$ & $0.06$ \\ 
        ...10 & $0.198$ & $1.00$ & $0.07$ & $1.00$ & $0.06$ & $1.00$ & $0.06$ & $1.00$ & $0.06$ \\ 
        ...11 & $0.318$ & $1.00$ & $0.06$ & $0.99$ & $0.05$ & $1.00$ & $0.05$ & $1.00$ & $0.05$ \\
        \hline
        \hline
        & \textbf{Average:} & 0.839 & 0.049 & 0.896 & 0.041 & 0.945 & 0.038 & 0.969 & 0.038\\
         \hline
         \hline
          \multicolumn{10}{c}{\footnotesize The positions of the variables are, in order, Citizenship, Age, Race, Sex, Income.}\\
          \multicolumn{10}{c}{\footnotesize Thus, cell ...11 represents females (\textit{SEX = 1}) with income above the federal poverty line (\textit{INC=1}).}\\
          \hline
    \end{tabular}
    \label{tab:repeatedSampling}
\end{table}

Table~\ref{tab:repeatedSampling} displays the 
coverage probabilities and average interval length for each marginal probability.  
In general, the credible intervals are close to or exceed the nominal coverage rate when $\epsilon = \infty$, confirming the usefulness of the latent class approach absent privacy concerns.  The coverage rates stay high for $\epsilon =1$, with a few coverage rates for small probabilities dipping to around 70\%.  The coverage rates degrade noticeably when $\epsilon = 0.25$, especially for the small probabilities.  Overall, we notice that 
coverage probabilities for cells corresponding to statistically insignificant terms in the log-linear model typically perform worse than other counts (e.g., cells \textit{0..0.} and \textit{1..1.}). This is because the noise has greater impact on small counts, and in a sample of size 10,000 many of these low probability cells have very few sampled cases.
We  conjecture that the undercoverage also relates, in part, to the sample size in our simulation design.
In fact, we repeated the simulation using samples of 100,000 records and found coverage rates at or exceeding the nominal rates for all cell probabilities and all four values of $\epsilon$.

\subsubsection{Comparisons to Other Methods}

\begin{figure}
	\centering
	\includegraphics[width=4.25in]{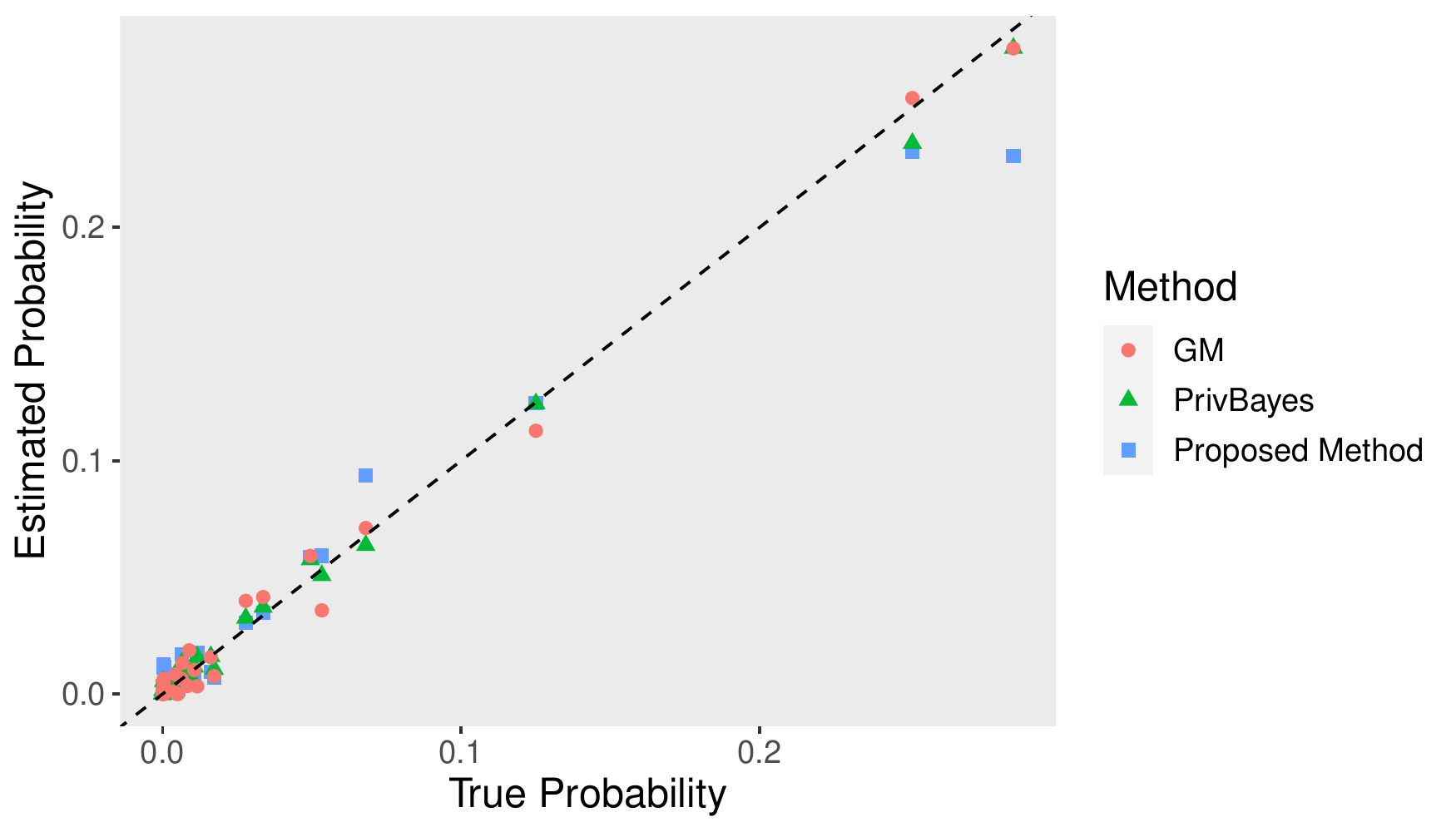}\\
	\includegraphics[width=4.25in]{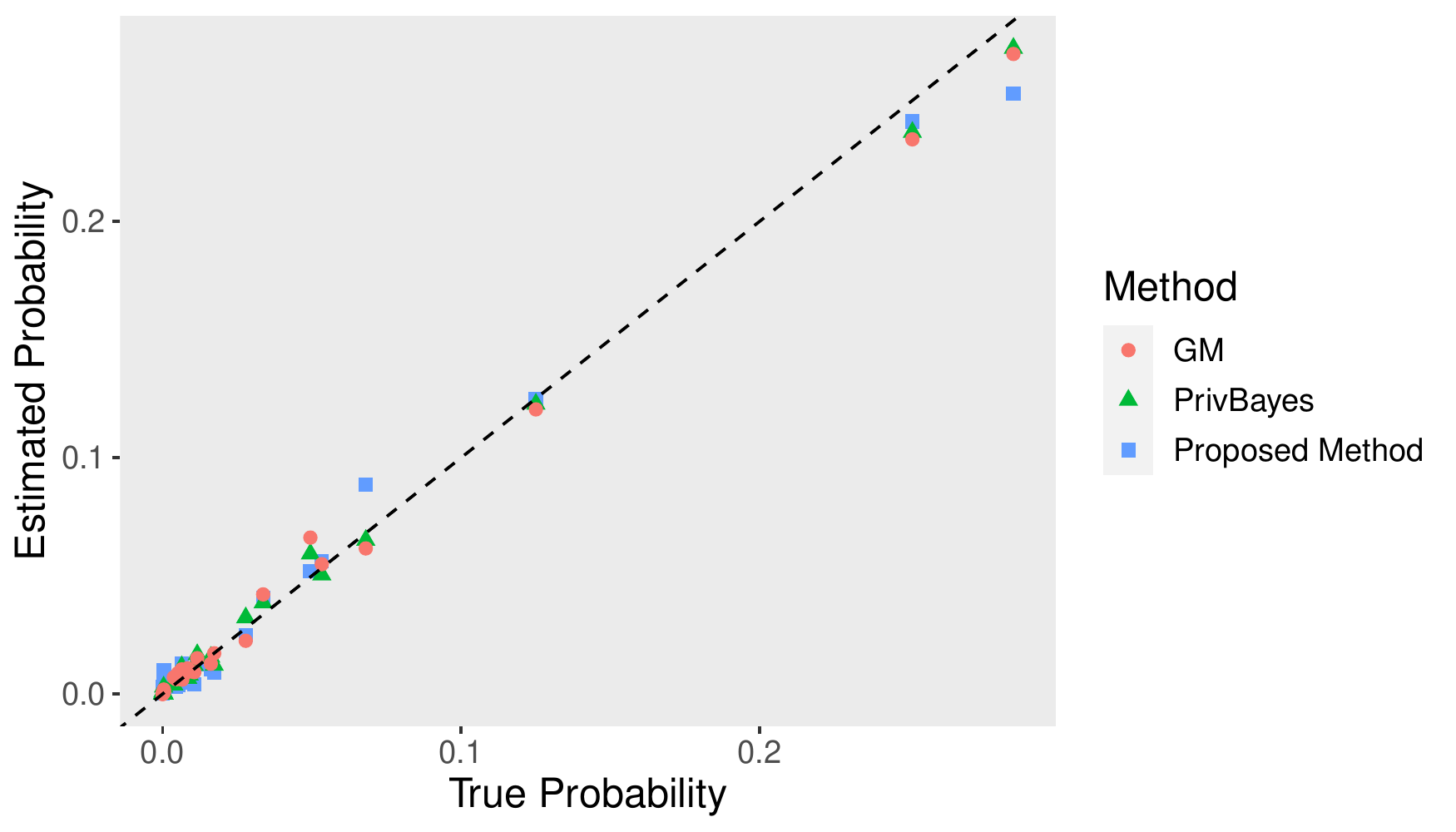}\\
	\includegraphics[width=4.25in]{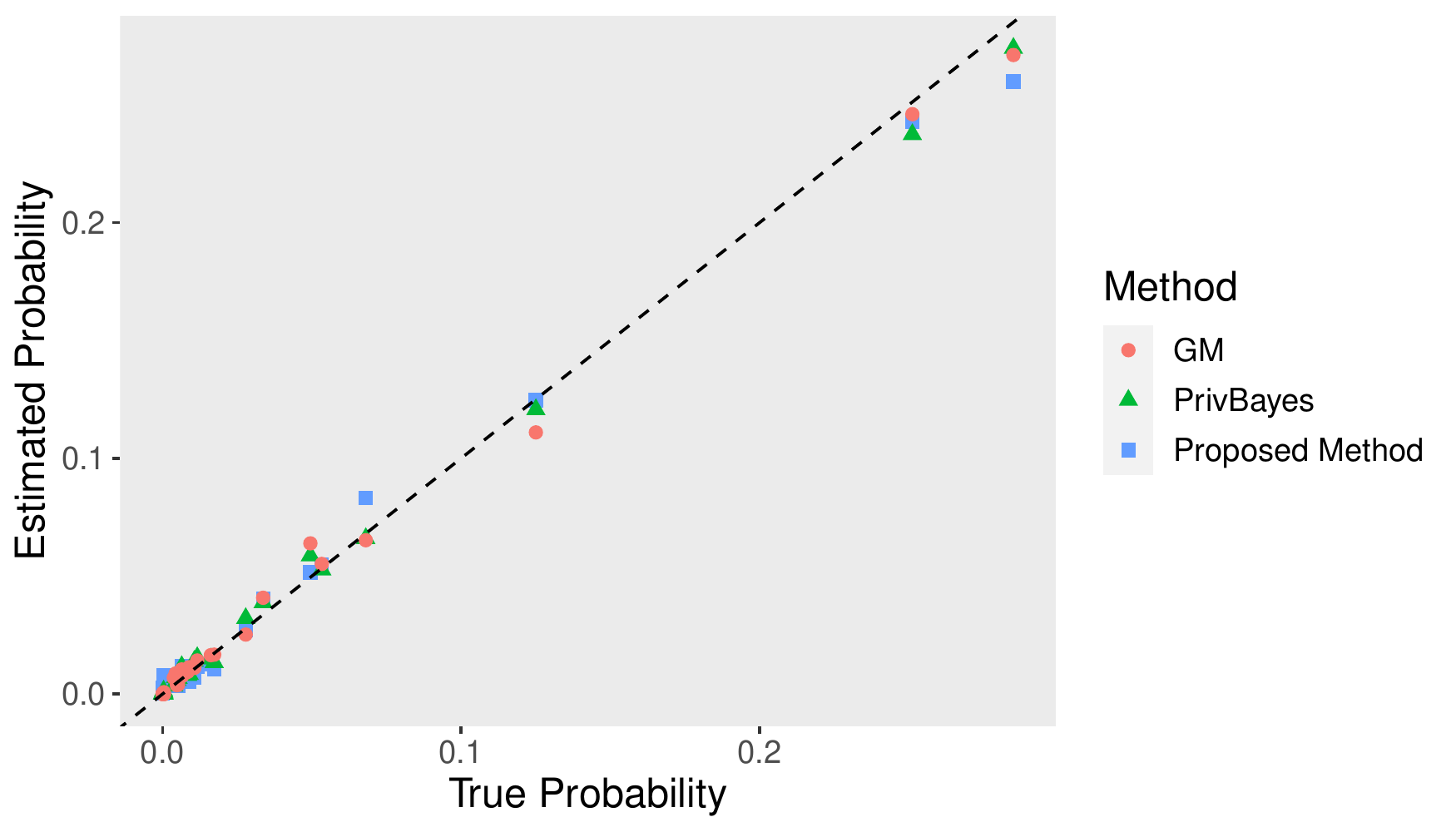}\\
	\caption{Plots of true full table probabilities versus estimated full table probabilities for $\epsilon \in \{0.25, 0.5, 1.0$\}. Comparison of latent class modeling approach with a graphical models-based approach and PrivBayes for $\epsilon = 0.25$ (top), $0.50$ (middle), and $1.00$ (bottom). For the proposed approach, estimated full cell probabilities denote the mean of the corresponding posterior distribution.}
	\label{fig:compare}
\end{figure}

We next compare the performance of our approach to PrivBayes and the graphical models approach. 
We ensure that all methods satisfy $\epsilon$-DP with the same $\epsilon$ and use consistent definitions of sensitivity. 
For the graphical models approach, we use all two way marginals as inputs. For PrivBayes, we set the maximum degree of the Bayesian network to two. Results are shown in Figure~\ref{fig:compare} for one run of each method. Results for other runs are similar. In general, all approaches return reasonable estimates for the marginal probabilities. Again, quality is dependent of the level of privacy. For smaller values of $\epsilon$, variability is most pronounced. While all methods perform extremely accurately for $\epsilon = 1.00$, one benefit of the latent class modeling approach is that it generates posterior inferences that account for the measurement error introduced by the privacy protection.

PrivBayes and the graphical models approach, rather than competitors, can complement our approach. For example, one can use PrivBayes to determine which marginal tables should be released and how to approximate the full table. Then, one could take the noisy and incoherent marginal tables obtained through PrivBayes and apply the graphical models approach to obtain a coherent version of these tables.  Thus, PrivBayes and graphical models techniques may be used to select which marginal tables to consider in our proposed latent class approach and to provide a starting point for $(M_1,\ldots,M_T)$ in the MCMC algorithm. We discuss some of these complementary uses in the next section.

\section{Application to the National Long-Term Care Survey}
\label{sec:NLTCS}

In this section we illustrate the latent class approach on a large-scale dataset from the National Long Term Care Survey (NLTCS).

\subsection{Data Description}
\label{sec:nltcs_data}

The NLTCS is a longitudinal survey sponsored by the U.S.\ National Institute on Aging\footnote{The NLTCS (National Long Term Care Study) is sponsored by the
National Institute of Aging and was conducted by the Duke University Center for
Demographic Studies under Grant No. U01-AG007198.}. First developed in 1982, the survey selects participants from Medicare enrollment files. All participants are over the age of 65, and new participants are added to the survey every five years \citep{nltcs}. Access to the data is restricted but can be obtained through an approved protocol \citep{nltcs-ispcr}.
For this analysis, we restricted ourselves to the 2004 wave of the NLTCS. In addition, we focused on sixteen variables representing markers of daily living as defined in Table~\ref{tab:nltcs-vars}. Participants with missing or unknown values for any of these variables were dropped, resulting in a $2^{16}$ contingency table comprised of $n = 15,636$ observations.

\begin{table}[ht]
    \caption{Data description for the NLTCS variables. Data was collected from the 2004 wave of the NLTCS. All variables had two levels: ``yes'' and ``no''.}
    \centering 
        \begin{tabular}{l l  l}
            \textbf{Variable} & \textbf{Description} & \textbf{Counts by Level} \\
            \hline
            SCN\_15\_A\_Y04 & Problem eating by self & ``yes'': 416; ``no'': 15,220\\ 
            SCN\_15\_B\_Y04 & Problem getting in/out of bed by self& ``yes'': 862; ``no'': 14,774\\
            SCN\_15\_C\_Y04 &Problem getting in/out of chairs by self& ``yes'': 1,054; ``no'': 14,582\\
            SCN\_15\_D\_Y04 &Problem walking inside without help& ``yes'': 1,763; ``no'': 13,873\\
            SCN\_15\_E\_Y04&Problem going outside without help& ``yes'': 2,435; ``no'': 13,201\\
            SCN\_15\_F\_Y04 & Problem dressing without help& ``yes'': 929; ``no'':  14,707\\
            SCN\_15\_G\_Y04 & Problem bathing without help&``yes'': 1,358; ``no'': 14,278\\
            SCN\_15\_H\_Y04 & Problem going to the bathroom without help&``yes'': 842; ``no'': 14,794\\
            SCN\_15\_I\_Y04 & Incontinence& ``yes'': 1,355; ``no'': 14,281\\
            SCN\_17\_A\_Y04 &Prepare meals without help& ``yes'': 13,670; ``no'': 1,966\\
            SCN\_17\_B\_Y04 &Do laundry without help& ``yes'': 13,465; ``no'': 2,171\\
            SCN\_17\_C\_Y04 &Light housework without help& ``yes'': 13,853; ``no'': 1,783\\
            SCN\_17\_D\_Y04 &Shop for groceries without help& ``yes'': 12,678; ``no'': 2,958\\
            SCN\_17\_E\_Y04 &Manage money without help& ``yes'': 13,737; ``no'': 1,899\\
            SCN\_17\_F\_Y04 &Take medicine without help& ``yes'': 14,031; ``no'': 1,605\\
            SCN\_17\_G\_Y04 &Make phone calls without help& ``yes'': 14,251; ``no'': 1,385\\
            \hline
        \end{tabular}
    \label{tab:nltcs-vars}
\end{table}

\subsection{Implementation Details}
\label{sec:nltcs_imp}

 The 16 binary variables make ${16 \choose 2} = 120$ two-way marginal tables and $120 \times (2 \times 2) = 480$ marginal counts. We selected important margins using the first half of the PrivBayes routine using a quarter of our privacy budget. While this allocates some of our privacy budget, it greatly reduces the number of marginals used in estimating model parameters. This resulted in 29 two-way margins. All counts were used for each pair, corresponding to $29   \times (2 \times 2) = 116$ marginal counts.

As before, we used a Metropolis-within-Gibbs algorithm to sample from the posterior distribution of $\boldsymbol{\pi_{k}}$, $\boldsymbol{\psi_{k}}$ and $(M_1,\ldots,M_k)$. To avoid overparametrization of the model, we set the number of latent classes to seven, ensuring that the number of estimated parameters (118) is roughly equal to the number of input marginals (116). We follow a similar implementation strategy as our ACS example; see Supplementary Section 2 for details. We run the chain for 12,000 iterations with the first 2,000 discarded as burn-in. We use four values of $\epsilon$: $0.5$, $1.0$, $2.0$, and $5.0$.  We estimate the two-way marginal probabilities at each iteration using $\boldsymbol{\pi_k}$ and $\boldsymbol{\psi_k}$.

\subsection{Results}
\label{sec:nltcs_results}

\begin{figure}
        \centering
        \includegraphics[width=3.5in]{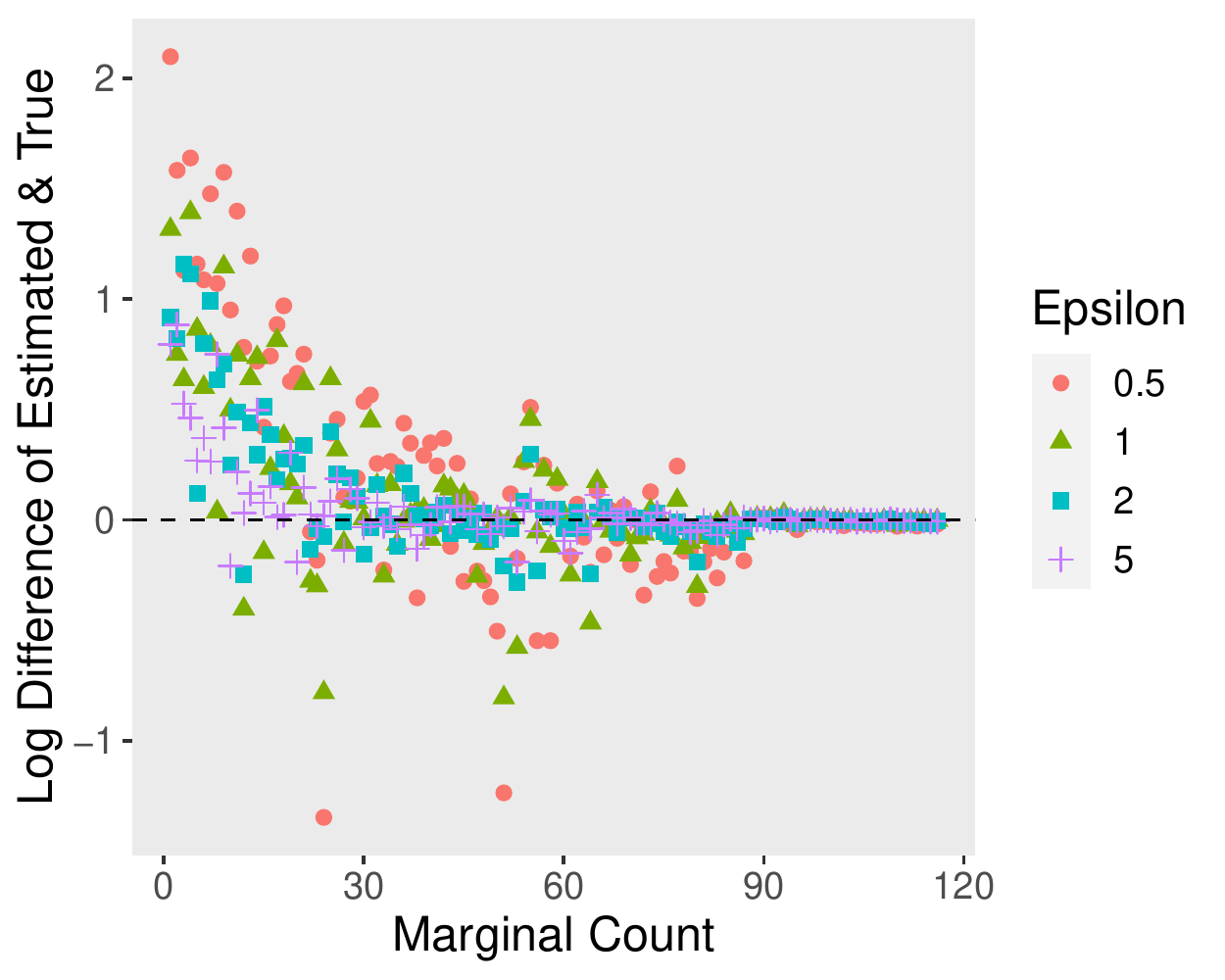}
        \caption{Plot of average log distance by marginal count for $\epsilon \in \{0.5, 1.0, 2.0, 5.0$\}. A log distance of zero indicates that the true and estimated probabilities are identical. Values greater than zero imply that the estimated probability is larger than the truth. Marginal counts are arranged in ascending order with the smallest counts on the left.}
        \label{fig:nltcs-results}
\end{figure}

For each of the 116 marginal counts, we compare the estimated and original (non-noisy) two-way probabilities. Figure~\ref{fig:nltcs-results} displays the log ratio (or difference e.g., $\log(P_{est}/P_{true})$) by $\epsilon$. The plotted marginal counts are ranked in order according to the true count with the smallest counts being on the left.  The average log difference across all counts ranged from 0.173 for $\epsilon = 0.5$ to 0.050 for $\epsilon = 5$. The estimates with $\epsilon = 0.5$ are improved over using the noisy marginal counts directly: the noisy marginal counts had an average log difference of -0.297 for $\epsilon = 0.5$.  However, when $\epsilon = 5$, the results from the noisy count are better, having an average log difference of 0.002.

Two observations are readily apparent from Figure~\ref{fig:nltcs-results}. First, log differences appear to decrease as a function of $\epsilon$ with higher values corresponding to greater similarity between the estimated and true marginal probabilities. Second, log differences appear to be greater for smaller marginal probabilities regardless of $\epsilon$.\footnote{In the true data, the smallest marginal probability is roughly 0.5\%, and about half of all marginal probabilities are smaller than 5\%.}
This result is somewhat unsurprising when considering the structure of our model: the variance of the binomial distribution directly corresponds to the probability. In fact, at the sample size of our data, the variance at the minimum marginal probability is 46 times smaller than the corresponding variance at 25\%. This effect is not as extreme at larger probabilities as the maximum marginal probability is approximately 92.5\%.

Furthermore, the vast majority of the log differences at these lower counts are positive, indicating that the estimated marginal probability is greater than the true marginal probability. We believe that this result is amplified by the inherent non-negativity of the counts and sampling from the simplex.

The main benefit of the latent class modeling approach is the capability to facilitate posterior inferences and to generate synthetic data. For the former, we computed 95\% credible intervals for each marginal probability by selecting the 2.5$^\mathrm{th}$ and 97.5$^\mathrm{th}$ quantiles of the posterior draws. For the latter, we randomly selected 20 iterates of the MCMC chain, and used each set of parameter draws to sample 15,636 synthetic records.  This creates 15,636 synthetic records that can summed to estimate probabilities in the full table.  We computed the marginal probabilities in each synthetic data set and combined the estimates using the approach in \cite{reiter:2003}, resulting in 95\% confidence intervals. Tables \ref{tab:nltcsExampsPostEst} and \ref{tab:nltcsExampsSynData} present results for three randomly selected marginal tables for the credible intervals and synthetic data inferences, respectively.  The posterior intervals and synthetic data intervals, which are quite similar, typically capture the true non-noisy estimate of the marginal probabilities, especially for larger values of $\epsilon$.

\begin{table}[!htbp] \centering 
      \caption{Posterior intervals for selected marginal counts in the NLTCS example.  Results also include the noisy counts input to the latent class model.} 
    \begin{tabular}{l l l l l l l l l}
             &  & &
        \multicolumn{2}{c}{\boldmath{$\epsilon$ = 0.5}}& \multicolumn{2}{c}{\boldmath{$\epsilon$ = 1.0}} & \multicolumn{2}{c}{\boldmath{$\epsilon$ = 2.0}}\\
        \cmidrule{4-9}
         &  & Pop. & \textit{Noisy} & & \textit{Noisy} &  & \textit{Noisy} & \\
         Variable& Cell & Est. & \textit{Est.} & \textit{95\% CI} & \textit{Est.} & \textit{95\% CI} & \textit{Est.} & \textit{95\% CI}\\
        \hline
        \hline
        & 00 & 0.041 & 0.033 & (0.053, 0.065) & 0.041& (0.044, 0.052) & 0.041& (0.040, 0.048)\\ 
        SCN\_15\_E & 01 & 0.114 & 0.112 & (0.093, 0.107) &0.117 & (0.106, 0.113) &0.114 & (0.107, 0.116)\\
        SCN\_17\_D & 10 & 0.769 & 0.776& (0.762, 0.783) & 0.770& (0.764, 0.774) & 0.768& (0.768, 0.780)\\
        & 11 & 0.075 & 0.076& (0.062, 0.080) &0.075 & (0.070, 0.077) &0.076 & (0.065, 0.076)\\
        \hline       
        & 00 & 0.853 &0.846 & (0.839, 0.851) & 0.849& (0.842, 0.849) & 0.854& (0.853, 0.858)\\ 
        SCN\_17\_B & 01 & 0.008 &0.007 & (0.016, 0.033) &0.010 & (0.013, 0.028) &0.008 & (0.007, 0.011)\\
        SCN\_17\_C & 10 & 0.033 & 0.039 & (0.032, 0.059) & 0.033& (0.035, 0.043) & 0.033& (0.030, 0.034)\\
        & 11 & 0.106 & 0.102& (0.067, 0.102) &0.107 & (0.089, 0.102) & 0.107& (0.102, 0.106)\\
        \hline        
        & 00 & 0.021 & 0.018& (0.013, 0.025) &0.021 & (0.011, 0.019) &0.021 & (0.015, 0.021)\\ 
        SCN\_15\_C & 01 & 0.046 & 0.046& (0.024, 0.044) &0.036 & (0.034, 0.045) & 0.046& (0.045, 0.050)\\
        SCN\_17\_A & 10 & 0.853 &0.864 & (0.846, 0.861) &0.854 & (0.855,0.866) & 0.852& (0.853, 0.859)\\
        & 11 & 0.080 &0.084 & (0.079, 0.105) & 0.085& (0.079, 0.093) &0.080 & (0.076, 0.081)\\
         \hline
         \hline
    \end{tabular}
    \label{tab:nltcsExampsPostEst}
\end{table} 

\begin{table}[!htbp] \centering 
      \caption{Interval estimates based on 20 synthetic data sets for selected marginal counts in the NLTCS example. Point estimates, standard errors, and multipliers were computed using the combining rules of \cite{reiter:2003}.}
    \begin{tabular}{l l l l l l l l l}
             &  & &
        \multicolumn{1}{c}{\boldmath{$\epsilon$ = 0.5}}& \multicolumn{1}{c}{\boldmath{$\epsilon$ = 1.0}} & \multicolumn{1}{c}{\boldmath{$\epsilon$ = 2.0}}\\
        \cmidrule{4-9}
         &  & Pop. & \textit{Point} & & \textit{Point} &  & \textit{Point} & \\
         Variable& Cell & Est. & \textit{Est.} & \textit{95\% CI} & \textit{Est.} & \textit{95\% CI} & \textit{Est.} & \textit{95\% CI}\\
        \hline
        \hline
        & 00 & 0.041 &  0.057 & (0.053, 0.061) & 0.044 & (0.041, 0.048) & 0.044 & (0.04, 0.047) \\ 
        SCN\_15\_E & 01 & 0.114 & 0.08 & (0.076, 0.085) & 0.091 & (0.086, 0.096) & 0.107 & (0.102, 0.112) \\ 
        SCN\_17\_D & 10 & 0.769 & 0.806 & (0.798, 0.815) & 0.793 & (0.786, 0.8) & 0.779 & (0.772, 0.786) \\ 
        & 11 & 0.075 & 0.056 & (0.05, 0.061) & 0.072 & (0.067, 0.076) & 0.07 & (0.066, 0.074) \\ 
        \hline       
        & 00 & 0.853 & 0.879 & (0.873, 0.886) & 0.869 & (0.864, 0.875) & 0.86 & (0.854, 0.865) \\ 
        SCN\_17\_B & 01 & 0.008 & 0.017 & (0.014, 0.019) & 0.012 & (0.01, 0.015) & 0.008 & (0.007, 0.01) \\ 
        SCN\_17\_C & 10 & 0.033 & 0.034 & (0.03, 0.039) & 0.03 & (0.027, 0.033) & 0.03 & (0.028, 0.033) \\ 
        & 11 & 0.106 & 0.07 & (0.061, 0.078) & 0.088 & (0.083, 0.093) & 0.102 & (0.097, 0.106) \\ 
        \hline        
        & 00 & 0.021 &  0.016 & (0.013, 0.018) & 0.011 & (0.008, 0.013) & 0.017 & (0.015, 0.019) \\ 
        SCN\_15\_C & 01 & 0.046 & 0.023 & (0.018, 0.028) & 0.032 & (0.029, 0.036) & 0.046 & (0.042, 0.049) \\ 
        SCN\_17\_A & 10 & 0.853 & 0.886 & (0.88, 0.892) & 0.882 & (0.876, 0.887) & 0.86 & (0.855, 0.866) \\ 
        & 11 & 0.080 &  0.076 & (0.07, 0.081) & 0.075 & (0.071, 0.08) & 0.077 & (0.073, 0.081) \\ 
         \hline
         \hline
    \end{tabular}
    \label{tab:nltcsExampsSynData}
\end{table}

\section{Concluding Remarks}
\label{sec:conclusions}

We present a novel method for posterior inferences and  creating differentially private synthetic data for contingency tables based on marginal counts. The simulation results indicate that the latent class modeling approach can offer reasonably accurate estimates of the counts used as inputs to construct the underlying tables (in our example, 2-way marginals). While it also did reasonably well in preserving counts from the full table in our illustrative data, analysts generally should not expect this to be the case.  Rather than as a mechanism for generating data coherent with the full table probabilities, we view the approach as a means to make posterior inferences and generate synthetic data that are coherent with the marginal counts used as inputs. Thus, we advise agencies to tell data users what counts are used as inputs, so that data users do not expect the synthetic data to be accurate for other counts.

More generally, our approach is to define summary statistics as functions of model parameters, and use composite likelihood approximations to estimate those parameters. This general strategy can be extended to more complex data structures.  For example, and as illustrated in the online supplement, the strategy could be used to generate synthetic data for people nested within households, using the latent class model of  \cite{hu2018dirichlet} as the underlying model. Developing methods for practical implementation in such complex models is an area for future research.

We have shown in the NLTCS example that not all marginals are required to estimate our model.  We leave it to future research to study methods for selecting the marginals, and, similarly, methods for allocating the privacy budget across these marginals. For both of our examples, we equally allocated the privacy budget. However, equal allocation may not be optimal in all problems.

\section*{Acknowledgment}

This work was supported by the Pennsylvania State University, Duke University, ACI under award 1443014, and the National Science Foundation under an IGERT award \#DGE-1144860, Big Data Social Science and awards SES-1534433 and SES-1131897.

\bibliography{references}

\begin{thebibliography}{77}
\providecommand{\natexlab}[1]{#1}
\providecommand{\url}[1]{\texttt{#1}}
\expandafter\ifx\csname urlstyle\endcsname\relax
  \providecommand{\doi}[1]{doi: #1}\else
  \providecommand{\doi}{doi: \begingroup \urlstyle{rm}\Url}\fi

\bibitem[Abowd(2018)]{abowd2018us}
J.~M. Abowd.
\newblock The {US} {C}ensus {B}ureau adopts differential privacy.
\newblock In \emph{Proceedings of the 24th ACM SIGKDD International Conference
  on Knowledge Discovery \& Data Mining}, pages 2867--2867. ACM, 2018.
\newblock \doi{10.1145/3219819.3226070}.

\bibitem[Akande et~al.(2019{\natexlab{a}})Akande, Barrientos, and
  Reiter]{akande2019simultaneous}
O.~Akande, A.~Barrientos, and J.~P. Reiter.
\newblock Simultaneous edit and imputation for household data with structural
  zeros.
\newblock \emph{Journal of Survey Statistics and Methodology}, 7\penalty0
  (4):\penalty0 498--519, 2019{\natexlab{a}}.
\newblock \doi{10.1093/jssam/smy022}.

\bibitem[Akande et~al.(2019{\natexlab{b}})Akande, Reiter, and
  Barrientos]{akande2017multiple}
O.~Akande, J.~Reiter, and A.~F. Barrientos.
\newblock Multiple imputation of missing values in household data with
  structural zeros.
\newblock \emph{Survey Methodology}, 2:\penalty0 498--519, 2019{\natexlab{b}}.
\newblock URL \url{https://hdl.handle.net/10161/17537}.

\bibitem[Awan and Slavkovic(2021)]{awanslav2021}
J.~Awan and A.~Slavkovic.
\newblock Structure and sensitivity in differential privacy: Comparing k-norm
  mechanisms.
\newblock \emph{Journal of the American Statistical Association}, 116\penalty0
  (534):\penalty0 935--954, 2021.
\newblock \doi{10.1080/01621459.2020.1773831}.
\newblock URL \url{https://doi.org/10.1080/01621459.2020.1773831}.

\bibitem[Bao et~al.(2021)Bao, Xiao, Zhao, Zhang, and Ding]{bao2021synthetic}
E.~Bao, X.~Xiao, J.~Zhao, D.~Zhang, and B.~Ding.
\newblock Synthetic data generation with differential privacy via bayesian
  networks.
\newblock \emph{Journal of Privacy and Confidentiality}, 11\penalty0 (3), 2021.

\bibitem[Barak et~al.(2007)Barak, Chaudhuri, Dwork, Kale, McSherry, and
  Talwar]{barak2007privacy}
B.~Barak, K.~Chaudhuri, C.~Dwork, S.~Kale, F.~McSherry, and K.~Talwar.
\newblock Privacy, accuracy, and consistency too: a holistic solution to
  contingency table release.
\newblock In \emph{Proceedings of the twenty-sixth ACM SIGMOD-SIGACT-SIGART
  symposium on Principles of database systems}, pages 273--282. ACM, 2007.
\newblock \doi{10.1145/1265530.1265569}.

\bibitem[Barrientos et~al.(2018)Barrientos, Bolton, Balmat, Reiter,
  de~Figueiredo, Machanavajjhala, Chen, Kneifel, DeLong,
  et~al.]{barrientos2018providing}
A.~F. Barrientos, A.~Bolton, T.~Balmat, J.~P. Reiter, J.~M. de~Figueiredo,
  A.~Machanavajjhala, Y.~Chen, C.~Kneifel, M.~DeLong, et~al.
\newblock Providing access to confidential research data through synthesis and
  verification: An application to data on employees of the {US} {F}ederal
  {G}overnment.
\newblock \emph{The Annals of Applied Statistics}, 12\penalty0 (2):\penalty0
  1124--1156, 2018.
\newblock \doi{10.1214/18-AOAS1194}.

\bibitem[Benedetto et~al.(2013)Benedetto, Stinson, and Abowd]{SIPP}
G.~Benedetto, M.~Stinson, and J.~M. Abowd.
\newblock The creation and use of the {SIPP} {Synthetic} {Beta}.
\newblock Mimeo, U.S. Census Bureau, Apr. 2013.
\newblock URL \url{http://hdl.handle.net/1813/43924}.

\bibitem[Bowen and Liu(2016)]{bowen2016comparative}
C.~M. Bowen and F.~Liu.
\newblock Comparative study of differentially private data synthesis methods.
\newblock \emph{arXiv preprint arXiv:1602.01063}, 2016.

\bibitem[Bowen and Snoke(2019)]{bowen2019comparative}
C.~M. Bowen and J.~Snoke.
\newblock Comparative study of differentially private synthetic data algorithms
  from the nist pscr differential privacy synthetic data challenge.
\newblock \emph{arXiv preprint arXiv:1911.12704}, 2019.

\bibitem[Caiola and Reiter(2010)]{caiola2010random}
G.~Caiola and J.~P. Reiter.
\newblock Random forests for generating partially synthetic, categorical data.
\newblock \emph{Transaction on Data Privacy}, 3:\penalty0 27--42, 2010.
\newblock \doi{10.5555/1747335.1747337}.

\bibitem[Charest(2012)]{charest2012empirical}
A.-S. Charest.
\newblock Empirical evaluation of statistical inference from
  differentially-private contingency tables.
\newblock In \emph{International Conference on Privacy in Statistical
  Databases}, pages 257--272. Springer, 2012.
\newblock \doi{10.1007/978-3-642-33627-0_20}.

\bibitem[Chaudhuri et~al.(2013)Chaudhuri, Sarwate, and
  Sinha]{chaudhuri2013near}
K.~Chaudhuri, A.~D. Sarwate, and K.~Sinha.
\newblock A near-optimal algorithm for differentially-private principal
  components.
\newblock \emph{Journal of Machine Learning Research}, 14, 2013.

\bibitem[Drechsler(2009)]{drechsler2009synthetic}
J.~Drechsler.
\newblock Synthetic datasets for the {German} {IAB} {E}stablishment {P}anel.
\newblock Invited Paper WP.10, Joint UNECE/Eurostat work session on statistical
  data confidentiality, 2009.
\newblock URL
  \url{http://www.unece.org/fileadmin/DAM/stats/documents/ece/ces/ge.46/2009/wp.10.e.pdf}.

\bibitem[Drechsler(2011)]{drechsler2011synthetic}
J.~Drechsler.
\newblock \emph{Synthetic datasets for statistical disclosure control: {T}heory
  and implementation}, volume 201 of \emph{Lecture Notes in Statistics}.
\newblock Springer Science \& Business Media, 2011.
\newblock ISBN 9781461403258.
\newblock \doi{10.1007/978-1-4614-0326-5}.

\bibitem[Dunson and Xing(2009)]{dunson2009nonparametric}
D.~B. Dunson and C.~Xing.
\newblock Nonparametric {B}ayes modeling of multivariate categorical data.
\newblock \emph{Journal of the American Statistical Association}, 104\penalty0
  (487):\penalty0 1042--1051, 2009.
\newblock \doi{10.1198/jasa.2009.tm08439}.

\bibitem[Dwork et~al.(2006{\natexlab{a}})Dwork, Kenthapadi, McSherry, Mironov,
  and Naor]{dwork2006our}
C.~Dwork, K.~Kenthapadi, F.~McSherry, I.~Mironov, and M.~Naor.
\newblock Our data, ourselves: Privacy via distributed noise generation.
\newblock In \emph{Annual International Conference on the Theory and
  Applications of Cryptographic Techniques}, pages 486--503. Springer,
  2006{\natexlab{a}}.
\newblock \doi{10.1007/11761679_29}.

\bibitem[Dwork et~al.(2006{\natexlab{b}})Dwork, McSherry, Nissim, and
  Smith]{dwork2006calibrating}
C.~Dwork, F.~McSherry, K.~Nissim, and A.~Smith.
\newblock Calibrating noise to sensitivity in private data analysis.
\newblock In \emph{Theory of cryptography conference}, pages 265--284.
  Springer, 2006{\natexlab{b}}.
\newblock \doi{10.1007/978-3-540-32732-5_32}.

\bibitem[Dwork et~al.(2014)Dwork, Roth, et~al.]{dwork2014algorithmic}
C.~Dwork, A.~Roth, et~al.
\newblock The algorithmic foundations of differential privacy.
\newblock \emph{Foundations and Trends in Theoretical Computer Science},
  9\penalty0 (3--4):\penalty0 211--407, 2014.
\newblock \doi{10.1561/0400000042}.

\bibitem[Eugenio and Liu(2018)]{eugenio2018construction}
E.~C. Eugenio and F.~Liu.
\newblock Construction of microdata from a set of differentially private
  low-dimensional contingency tables through solving linear equations with
  tikhonov regularization.
\newblock \emph{arXiv preprint arXiv:1812.05671}, 2018.

\bibitem[Ferguson(1973)]{ferguson1973bayesian}
T.~S. Ferguson.
\newblock A bayesian analysis of some nonparametric problems.
\newblock \emph{The annals of statistics}, pages 209--230, 1973.
\newblock URL \url{https://www.jstor.org/stable/2958008}.

\bibitem[Fienberg and Slavkovic(2005)]{fienberg2005preserving}
S.~E. Fienberg and A.~B. Slavkovic.
\newblock Preserving the confidentiality of categorical statistical data bases
  when releasing information for association rules.
\newblock \emph{Data Mining and Knowledge Discovery}, 11\penalty0 (2):\penalty0
  155--180, 2005.

\bibitem[Fuller(2009)]{fuller2009measurement}
W.~A. Fuller.
\newblock \emph{Measurement Error Models}, volume 305.
\newblock John Wiley \& Sons, 2009.
\newblock \doi{10.1002/9780470316665}.

\bibitem[Garfinkel et~al.(2018)Garfinkel, Abowd, and
  Powazek]{garfinkel2018issues}
S.~L. Garfinkel, J.~M. Abowd, and S.~Powazek.
\newblock Issues encountered deploying differential privacy.
\newblock In \emph{Proceedings of the 2018 Workshop on Privacy in the
  Electronic Society}, pages 133--137. ACM, 2018.
\newblock \doi{10.1145/3267323.3268949}.

\bibitem[Gelman et~al.(2013)Gelman, Carlin, Stern, Dunson, Vehtari, and
  Rubin]{gelman2013bayesian}
A.~Gelman, J.~B. Carlin, H.~S. Stern, D.~B. Dunson, A.~Vehtari, and D.~B.
  Rubin.
\newblock \emph{Bayesian data analysis}.
\newblock CRC press, 2013.

\bibitem[Ghosh et~al.(2012)Ghosh, Roughgarden, and
  Sundararajan]{ghosh2012universally}
A.~Ghosh, T.~Roughgarden, and M.~Sundararajan.
\newblock Universally utility-maximizing privacy mechanisms.
\newblock \emph{SIAM Journal on Computing}, 41\penalty0 (6):\penalty0
  1673--1693, 2012.
\newblock \doi{10.1145/1536414.1536464}.

\bibitem[Gong(2019)]{gong2019exact}
R.~Gong.
\newblock Exact inference with approximate computation for differentially
  private data via perturbations.
\newblock \emph{arXiv preprint arXiv:1909.12237}, 2019.

\bibitem[Hardt et~al.(2012)Hardt, Ligett, and McSherry]{hardt2012simple}
M.~Hardt, K.~Ligett, and F.~McSherry.
\newblock A simple and practical algorithm for differentially private data
  release.
\newblock In \emph{Advances in Neural Information Processing Systems}, pages
  2339--2347, 2012.
\newblock \doi{10.5555/2999325.2999396}.

\bibitem[Hay et~al.(2010)Hay, Rastogi, Miklau, and Suciu]{hay2010boosting}
M.~Hay, V.~Rastogi, G.~Miklau, and D.~Suciu.
\newblock Boosting the accuracy of differentially private histograms through
  consistency.
\newblock \emph{Proceedings of the VLDB Endowment}, 3\penalty0 (1-2):\penalty0
  1021--1032, 2010.
\newblock \doi{10.14778/1920841.1920970}.

\bibitem[Hu et~al.(2018)Hu, Reiter, Wang, et~al.]{hu2018dirichlet}
J.~Hu, J.~P. Reiter, Q.~Wang, et~al.
\newblock Dirichlet process mixture models for modeling and generating
  synthetic versions of nested categorical data.
\newblock \emph{Bayesian Analysis}, 13\penalty0 (1):\penalty0 183--200, 2018.
\newblock \doi{10.1214/16-BA1047}.

\bibitem[Hundepool et~al.(2012)Hundepool, Domingo-Ferrer, Franconi, Giessing,
  Nordholt, Spicer, and De~Wolf]{hundepool}
A.~Hundepool, J.~Domingo-Ferrer, L.~Franconi, S.~Giessing, E.~S. Nordholt,
  K.~Spicer, and P.-P. De~Wolf.
\newblock \emph{Statistical disclosure control}.
\newblock John Wiley \& Sons, 2012.
\newblock \doi{10.1002/9781118348239}.

\bibitem[Inusah and Kozubowski(2006)]{inusah2006discrete}
S.~Inusah and T.~J. Kozubowski.
\newblock A discrete analogue of the {L}aplace distribution.
\newblock \emph{Journal of statistical planning and inference}, 136\penalty0
  (3):\penalty0 1090--1102, 2006.
\newblock \doi{10.1016/j.jspi.2004.08.014}.

\bibitem[Ishwaran and James(2001)]{ishwaran2001gibbs}
H.~Ishwaran and L.~F. James.
\newblock Gibbs sampling methods for stick-breaking priors.
\newblock \emph{Journal of the American Statistical Association}, 96\penalty0
  (453):\penalty0 161--173, 2001.
\newblock \doi{10.1198/016214501750332758}.

\bibitem[Karr et~al.(2006)Karr, Kohnen, Oganian, Reiter, and
  Sanil]{karr2006framework}
A.~F. Karr, C.~N. Kohnen, A.~Oganian, J.~P. Reiter, and A.~P. Sanil.
\newblock A framework for evaluating the utility of data altered to protect
  confidentiality.
\newblock \emph{The American Statistician}, 60\penalty0 (3):\penalty0 224--232,
  2006.
\newblock \doi{10.1198/000313006X124640}.

\bibitem[Karwa et~al.(2015)Karwa, Kifer, and Slavkovi{\'c}]{karwa2015private}
V.~Karwa, D.~Kifer, and A.~B. Slavkovi{\'c}.
\newblock Private posterior distributions from variational approximations.
\newblock \emph{arXiv preprint arXiv:1511.07896}, 2015.

\bibitem[Karwa et~al.(2017)Karwa, Krivitsky, and
  Slavkovi{\'c}]{karwa2017sharing}
V.~Karwa, P.~N. Krivitsky, and A.~B. Slavkovi{\'c}.
\newblock Sharing social network data: differentially private estimation of
  exponential family random-graph models.
\newblock \emph{Journal of the Royal Statistical Society Series C}, 66\penalty0
  (3):\penalty0 481--500, 2017.
\newblock \doi{10.1111/rssc.12185}.

\bibitem[Kinney et~al.(2011)Kinney, Reiter, Reznek, Miranda, Jarmin, and
  Abowd]{synlbd}
S.~K. Kinney, J.~P. Reiter, A.~P. Reznek, J.~Miranda, R.~S. Jarmin, and J.~M.
  Abowd.
\newblock Towards unrestricted public use business microdata: The synthetic
  {L}ongitudinal {B}usiness {D}atabase.
\newblock \emph{International Statistical Review}, 79\penalty0 (3):\penalty0
  362--384, 2011.
\newblock \doi{10.1111.j.1751-5823.2011.00153.x}.

\bibitem[Lee et~al.(2015)Lee, Wang, and Kifer]{lee2015maximum}
J.~Lee, Y.~Wang, and D.~Kifer.
\newblock Maximum likelihood postprocessing for differential privacy under
  consistency constraints.
\newblock In \emph{Proceedings of the 21th ACM SIGKDD International Conference
  on Knowledge Discovery and Data Mining}, pages 635--644, 2015.
\newblock \doi{10.1145/2783258.2783366}.

\bibitem[Li et~al.(2018)Li, Karwa, Slavkovi{\'c}, and Steorts]{li2018privacy}
B.~Li, V.~Karwa, A.~Slavkovi{\'c}, and R.~C. Steorts.
\newblock A privacy preserving algorithm to release sparse high-dimensional
  histograms.
\newblock \emph{Journal of Privacy and Confidentiality}, 8\penalty0 (1), 2018.
\newblock \doi{10.29012/jpc.657}.

\bibitem[Lindsay(1988)]{lindsay1988composite}
B.~G. Lindsay.
\newblock Composite likelihood methods.
\newblock \emph{Contemporary mathematics}, 80\penalty0 (1):\penalty0 221--239,
  1988.
\newblock \doi{10.1090/conm/080/999014}.

\bibitem[Little(1993)]{little1993statistical}
R.~J. Little.
\newblock Statistical analysis of masked data.
\newblock \emph{Journal of Official Statistics}, 9\penalty0 (2):\penalty0
  407--426, 1993.

\bibitem[Machanavajjhala et~al.(2008)Machanavajjhala, Kifer, Abowd, Gehrke, and
  Vilhuber]{Ashwin2008}
A.~Machanavajjhala, D.~Kifer, J.~M. Abowd, J.~Gehrke, and L.~Vilhuber.
\newblock Privacy: {T}heory meets practice on the map.
\newblock \emph{International Conference on Data Engineering (ICDE)}, pages
  277--286, 2008.
\newblock \doi{10.1109/ICDE.2008.4497436}.

\bibitem[Manrique-Vallier and Reiter(2014)]{manrique2014bayesian}
D.~Manrique-Vallier and J.~P. Reiter.
\newblock Bayesian estimation of discrete multivariate latent structure models
  with structural zeros.
\newblock \emph{Journal of Computational and Graphical Statistics}, 23\penalty0
  (4):\penalty0 1061--1079, 2014.
\newblock \doi{10.1080/10618600.2013.844700}.

\bibitem[Manton(2010)]{nltcs-ispcr}
K.~G. Manton.
\newblock National long-term care survey: 1982, 1984, 1989, 1994, 1999, and
  2004, 2010.
\newblock URL \url{https://doi.org/10.3886/ICPSR09681.v5}.

\bibitem[Mateo-Sanz et~al.(2004)Mateo-Sanz, Mart{\'\i}nez-Ballest{\'e}, and
  Domingo-Ferrer]{mateo2004fast}
J.~M. Mateo-Sanz, A.~Mart{\'\i}nez-Ballest{\'e}, and J.~Domingo-Ferrer.
\newblock Fast generation of accurate synthetic microdata.
\newblock In \emph{International Workshop on Privacy in Statistical Databases},
  pages 298--306. Springer, 2004.
\newblock \doi{10.1007/978-3-540-25955-8_24}.

\bibitem[McKenna et~al.(2019)McKenna, Sheldon, and
  Miklau]{mckenna2019graphical}
R.~McKenna, D.~Sheldon, and G.~Miklau.
\newblock Graphical-model based estimation and inference for differential
  privacy.
\newblock \emph{arXiv preprint arXiv:1901.09136}, 2019.

\bibitem[McKenna et~al.(2021{\natexlab{a}})McKenna, Miklau, and
  Sheldon]{McKenna_Miklau_Sheldon_2021}
R.~McKenna, G.~Miklau, and D.~Sheldon.
\newblock Winning the {NIST} contest: A scalable and general approach to
  differentially private synthetic data.
\newblock 11, 2021{\natexlab{a}}.
\newblock URL
  \url{https://journalprivacyconfidentiality.org/index.php/jpc/article/view/778}.

\bibitem[McKenna et~al.(2021{\natexlab{b}})McKenna, Miklau, and
  Sheldon]{mckenna2021winning}
R.~McKenna, G.~Miklau, and D.~Sheldon.
\newblock Winning the nist contest: A scalable and general approach to
  differentially private synthetic data.
\newblock \emph{Journal of Privacy and Confidentiality}, 11\penalty0 (3),
  2021{\natexlab{b}}.

\bibitem[McSherry(2009)]{mcsherry2009privacy}
F.~D. McSherry.
\newblock Privacy integrated queries: an extensible platform for
  privacy-preserving data analysis.
\newblock In \emph{Proceedings of the 2009 ACM SIGMOD International Conference
  on Management of data}, pages 19--30. ACM, 2009.
\newblock \doi{10.1145/1559845.1559850}.

\bibitem[Mironov(2017)]{mironov2017renyi}
I.~Mironov.
\newblock R{\'e}nyi differential privacy.
\newblock In \emph{2017 IEEE 30th Computer Security Foundations Symposium
  (CSF)}, pages 263--275. IEEE, 2017.
\newblock \doi{10.1109/CSF.2017.11}.

\bibitem[{National Institute of Standards and Technology}(2021)]{nist2018}
{National Institute of Standards and Technology}.
\newblock 2018 differential privacy synthetic data challenge, 2021.
\newblock URL
  \url{https://www.nist.gov/ctl/pscr/open-innovation-prize-challenges/past-prize-challenges/2018-differential-privacy-synthetic}.

\bibitem[{National Institute on Aging}(2021)]{nltcs}
{National Institute on Aging}.
\newblock National long term care survey (nltcs), 2021.
\newblock URL
  \url{https://www.nia.nih.gov/research/resource/national-long-term-care-survey-nltcs}.

\bibitem[Park and Ghosh(2014)]{park2014pegs}
Y.~Park and J.~Ghosh.
\newblock Pegs: Perturbed gibbs samplers that generate privacy-compliant
  synthetic data.
\newblock \emph{Transactions on Data Privacy}, 7\penalty0 (3):\penalty0
  253--282, 2014.
\newblock \doi{10.5555/2870614.2870617}.

\bibitem[Ping et~al.(2017)Ping, Stoyanovich, and Howe]{ping2017datasynthesizer}
H.~Ping, J.~Stoyanovich, and B.~Howe.
\newblock Datasynthesizer: Privacy-preserving synthetic datasets.
\newblock In \emph{Proceedings of the 29th International Conference on
  Scientific and Statistical Database Management}, pages 1--5, 2017.
\newblock \doi{10.1145/3085504.3091117}.

\bibitem[Raab(2019)]{raab2019practical}
G.~Raab.
\newblock Practical experience with making synthetic data differentially
  private, 2019.
\newblock URL \url{https://simons.berkeley.edu/talks/tba-47}.

\bibitem[Reiter(2003)]{reiter:2003}
J.~P. Reiter.
\newblock Inference for partially synthetic, public use microdata sets.
\newblock \emph{Survey Methodology}, 29:\penalty0 181--188, 2003.

\bibitem[Reiter(2005{\natexlab{a}})]{reiter2005estimating}
J.~P. Reiter.
\newblock Estimating risks of identification disclosure in microdata.
\newblock \emph{Journal of the American Statistical Association}, 100\penalty0
  (472):\penalty0 1103--1112, 2005{\natexlab{a}}.
\newblock \doi{10.1198/016214505000000619}.

\bibitem[Reiter(2005{\natexlab{b}})]{reiter2005using}
J.~P. Reiter.
\newblock Using {CART} to generate partially synthetic public use microdata.
\newblock \emph{Journal of Official Statistics}, 21\penalty0 (3):\penalty0
  441--462, 2005{\natexlab{b}}.

\bibitem[Reiter et~al.(2014)Reiter, Wang, and Zhang]{reiter2014bayesian}
J.~P. Reiter, Q.~Wang, and B.~Zhang.
\newblock Bayesian estimation of disclosure risks for multiply imputed,
  synthetic data.
\newblock \emph{Journal of Privacy and Confidentiality}, 6\penalty0 (1), 2014.
\newblock \doi{10.29012/jpc.v6i1.635}.

\bibitem[Rinott et~al.(2018)Rinott, O’Keefe, Shlomo, Skinner,
  et~al.]{rinott2018confidentiality}
Y.~Rinott, C.~M. O’Keefe, N.~Shlomo, C.~Skinner, et~al.
\newblock Confidentiality and differential privacy in the dissemination of
  frequency tables.
\newblock \emph{Statistical Science}, 33\penalty0 (3):\penalty0 358--385, 2018.
\newblock URL \url{https://projecteuclid.org/euclid.ss/1534147228}.

\bibitem[Roy(2020)]{roy2020convergence}
V.~Roy.
\newblock Convergence diagnostics for markov chain monte carlo.
\newblock \emph{Annual Review of Statistics and Its Application}, 7:\penalty0
  387--412, 2020.

\bibitem[Rubin(1993)]{rubin1993discussion}
D.~B. Rubin.
\newblock Discussion: {S}tatistical disclosure limitation.
\newblock \emph{Journal of Official Statistics}, 9\penalty0 (2):\penalty0
  461--468, 1993.

\bibitem[Seeman et~al.(2020)Seeman, Slavkovic, and Reimherr]{seeman2020private}
J.~Seeman, A.~Slavkovic, and M.~Reimherr.
\newblock Private posterior inference consistent with public information: a
  case study in small area estimation from synthetic census data.
\newblock In \emph{International Workshop on Privacy in Statistical Databases}.
  Springer, 2020.
\newblock \doi{10.1007/978-3-030-57521-2_23}.

\bibitem[Si and Reiter(2013)]{si2013nonparametric}
Y.~Si and J.~P. Reiter.
\newblock Nonparametric bayesian multiple imputation for incomplete categorical
  variables in large-scale assessment surveys.
\newblock \emph{Journal of Educational and Behavioral Statistics}, 38\penalty0
  (5):\penalty0 499--521, 2013.
\newblock \doi{10.3102/1076998613480394}.

\bibitem[Slavkovi{\'c} et~al.(2015)Slavkovi{\'c}, Zhu, and
  Petrovi{\'c}]{slavzhupet2015}
A.~Slavkovi{\'c}, X.~Zhu, and S.~Petrovi{\'c}.
\newblock Fibers of multi-way contingency tables given conditionals: relation
  to marginals, cell bounds and markov bases.
\newblock \emph{Annals of the Institute of Statistical Mathematics},
  67\penalty0 (4):\penalty0 621--648, 2015.
\newblock \doi{10.1007/s10463-014-0471-z}.
\newblock URL \url{https://doi.org/10.1007/s10463-014-0471-z}.

\bibitem[Slavkovi{\'c} and Lee(2010)]{slavkovic2010synthetic}
A.~B. Slavkovi{\'c} and J.~Lee.
\newblock Synthetic two-way contingency tables that preserve conditional
  frequencies.
\newblock \emph{Statistical Methodology}, 7\penalty0 (3):\penalty0 225--239,
  2010.
\newblock \doi{10.1016/j.stamet.2009.11.002}.

\bibitem[SLS-DSU(2018)]{iCEM}
SLS-DSU.
\newblock {S}cottish {L}ongitudinal {S}tudy {D}evelopment \& {S}upport {U}nit,
  2018.
\newblock URL \url{https://sls.lscs.ac.uk/}.

\bibitem[Snoke and Slavkovi{\'c}(2018)]{snoke2018pmse}
J.~Snoke and A.~Slavkovi{\'c}.
\newblock p{MSE} mechanism: Differentially private synthetic data with maximal
  distributional similarity.
\newblock \emph{arXiv preprint arXiv:1805.09392}, 2018.

\bibitem[Snoke et~al.(2018)Snoke, Raab, Nowok, Dibben, and Slavkovic]{snoke}
J.~Snoke, G.~M. Raab, B.~Nowok, C.~Dibben, and A.~Slavkovic.
\newblock General and specific utility measures for synthetic data.
\newblock \emph{Journal of the Royal Statistical Society: Series A (Statistics
  in Society)}, 181\penalty0 (3):\penalty0 663--688, 2018.
\newblock \doi{10.1111/rssa.12358}.

\bibitem[Ullman and Vadhan(2020)]{ullman2020pcps}
J.~Ullman and S.~Vadhan.
\newblock Pcps and the hardness of generating synthetic data.
\newblock \emph{Journal of Cryptology}, 33\penalty0 (4):\penalty0 2078--2112,
  2020.

\bibitem[Varin et~al.(2011)Varin, Reid, and Firth]{varin2011overview}
C.~Varin, N.~Reid, and D.~Firth.
\newblock An overview of composite likelihood methods.
\newblock \emph{Statistica Sinica}, pages 5--42, 2011.
\newblock URL \url{https://www.jstor.org/stable/24309261}.

\bibitem[Walker(2013)]{walker2013bayesian}
S.~G. Walker.
\newblock Bayesian inference with misspecified models.
\newblock \emph{Journal of Statistical Planning and Inference}, 143\penalty0
  (10):\penalty0 1621--1633, 2013.
\newblock \doi{10.1016/j.jspi.2013.05.013}.

\bibitem[Woo and Slavkovi{\'c}(2012)]{woo2012logistic}
Y.~M.~J. Woo and A.~B. Slavkovi{\'c}.
\newblock Logistic regression with variables subject to post randomization
  method.
\newblock In \emph{International Conference on Privacy in Statistical
  Databases}, pages 116--130. Springer, 2012.
\newblock \doi{10.1007/978-3-642-33627-0_10}.

\bibitem[Woodcock and Benedetto(2009)]{woodcock2009distribution}
S.~D. Woodcock and G.~Benedetto.
\newblock Distribution-preserving statistical disclosure limitation.
\newblock \emph{Computational Statistics \& Data Analysis}, 53\penalty0
  (12):\penalty0 4228--4242, 2009.
\newblock \doi{10.1016/j.csda.2009.05.020}.

\bibitem[Yang et~al.(2012)Yang, Fienberg, and Rinaldo]{yang2012differential}
X.~Yang, S.~E. Fienberg, and A.~Rinaldo.
\newblock Differential privacy for protecting multi-dimensional contingency
  table data: Extensions and applications.
\newblock \emph{Journal of Privacy and Confidentiality}, 4\penalty0 (1), 2012.
\newblock \doi{10.29012/jpc.v4i1.613}.

\bibitem[Zhang et~al.(2012)Zhang, Zhang, Xiao, Yang, and
  Winslett]{zhang2012functional}
J.~Zhang, Z.~Zhang, X.~Xiao, Y.~Yang, and M.~Winslett.
\newblock Functional mechanism: regression analysis under differential privacy.
\newblock \emph{Proceedings of the VLDB Endowment}, 5\penalty0 (11):\penalty0
  1364--1375, 2012.
\newblock \doi{10.14778/2350229.2350253}.

\bibitem[Zhang et~al.(2017)Zhang, Cormode, Procopiuc, Srivastava, and
  Xiao]{zhang2017privbayes}
J.~Zhang, G.~Cormode, C.~M. Procopiuc, D.~Srivastava, and X.~Xiao.
\newblock Priv{B}ayes: Private data release via {B}ayesian networks.
\newblock \emph{ACM Transactions on Database Systems}, 42\penalty0
  (4):\penalty0 1--41, 2017.
\newblock \doi{10.1145/3134428}.

\end{thebibliography}


\begin{thebibliography}{7}
\providecommand{\natexlab}[1]{#1}
\providecommand{\url}[1]{\texttt{#1}}
\expandafter\ifx\csname urlstyle\endcsname\relax
  \providecommand{\doi}[1]{doi: #1}\else
  \providecommand{\doi}{doi: \begingroup \urlstyle{rm}\Url}\fi

\bibitem[Akande et~al.(2019)Akande, Barrientos, and
  Reiter]{akande2019simultaneous}
O.~Akande, A.~Barrientos, and J.~P. Reiter.
\newblock Simultaneous edit and imputation for household data with structural
  zeros.
\newblock \emph{Journal of Survey Statistics and Methodology}, 7\penalty0
  (4):\penalty0 498--519, 2019.
\newblock \doi{10.1093/jssam/smy022}.

\bibitem[Dunson and Xing(2009)]{dunson2009nonparametric}
D.~B. Dunson and C.~Xing.
\newblock Nonparametric {B}ayes modeling of multivariate categorical data.
\newblock \emph{Journal of the American Statistical Association}, 104\penalty0
  (487):\penalty0 1042--1051, 2009.
\newblock \doi{10.1198/jasa.2009.tm08439}.

\bibitem[Hu et~al.(2018)Hu, Reiter, Wang, et~al.]{hu2018dirichlet}
J.~Hu, J.~P. Reiter, Q.~Wang, et~al.
\newblock Dirichlet process mixture models for modeling and generating
  synthetic versions of nested categorical data.
\newblock \emph{Bayesian Analysis}, 13\penalty0 (1):\penalty0 183--200, 2018.
\newblock \doi{10.1214/16-BA1047}.

\bibitem[Ishwaran and James(2001)]{ishwaran2001gibbs}
H.~Ishwaran and L.~F. James.
\newblock Gibbs sampling methods for stick-breaking priors.
\newblock \emph{Journal of the American Statistical Association}, 96\penalty0
  (453):\penalty0 161--173, 2001.
\newblock \doi{10.1198/016214501750332758}.

\bibitem[Ishwaran and Zarepour(2000)]{ishwaran2000markov}
H.~Ishwaran and M.~Zarepour.
\newblock Markov chain monte carlo in approximate dirichlet and beta
  two-parameter process hierarchical models.
\newblock \emph{Biometrika}, 87\penalty0 (2):\penalty0 371--390, 2000.

\bibitem[Papadakis et~al.(2021)Papadakis, Tsagris, Dimitriadis, Fafalios,
  Tsamardinos, Fasiolo, Borboudakis, Burkardt, Zou, Lakiotaki, and
  Chatzipantsiou.]{RfastPackage}
M.~Papadakis, M.~Tsagris, M.~Dimitriadis, S.~Fafalios, I.~Tsamardinos,
  M.~Fasiolo, G.~Borboudakis, J.~Burkardt, C.~Zou, K.~Lakiotaki, and
  C.~Chatzipantsiou.
\newblock \emph{Rfast: A Collection of Efficient and Extremely Fast R
  Functions}, 2021.
\newblock URL \url{https://CRAN.R-project.org/package=Rfast}.
\newblock R package version 2.0.3.

\bibitem[{R Core Team}(2021)]{parallelPackage}
{R Core Team}.
\newblock \emph{R: A Language and Environment for Statistical Computing}.
\newblock R Foundation for Statistical Computing, Vienna, Austria, 2021.
\newblock URL \url{https://www.R-project.org/}.

\end{thebibliography}
\bibliographystyle{abbrvnat}

\end{document}


\title{Supplementary Materials for \textit{A Latent Class Modeling Approach for Generating Synthetic Data and Making Posterior Inferences from Differentially Private Counts}}
\author{Michelle Pistner Nixon, Andr\'es F. Barrientos, Jerome P. Reiter, and Aleksandra Slavkovi\'{c}}
\date{}
\maketitle

\section{Nested Data Example}
\label{sec:nestedApp}

In this section, we illustrate the latent class modeling strategy for nested categorical data, based on the model of \cite{hu2018dirichlet}.  

\subsection{Modeling strategy}

To facilitate understanding, we describe the method in the context of a subset of data from the 2012 American Community Survey (ACS). The ACS data have a nested structure: the data file has information on individuals nested within households. For simplicity, we restrict our focus to households of size three, i.e., a head of household and two other occupants. 
Thus, the data comprise information about the household head, each individual, relationships of the individuals to the household head (e.g., spouse, child), and about the household as a unit (e.g., is the house owned or rented).  Table~\ref{tab:nestedVars} includes a description of the variables.  We treat data for the household head as part of the household-level information. We treat data available for the two remaining family members as individual-level information.

Before describing the approach, we introduce some notation for ownership ($X_0$); the household head's gender and race $(X_1,X_2)$; the first other individual's gender, race, and relationship to household head $(X_{11},X_{12},X_{13})$; and, the second other individual's gender, race, and relationship to household head $(X_{21},X_{22},X_{23})$.

\begin{table}[ht]
    \footnotesize
    \caption{Variable description of nested ACS data. ``HH'' denotes head of household. See \cite{akande2019simultaneous} for remaining variable definitions.}
    \centering 
        \begin{tabular}{l  l}
            \textbf{Variable} & \textbf{Levels} \\
            \hline
            &\\
            \multicolumn{2}{c}{\underline{\textit{Household-level}}}\\
            &\\
            Household Index & Household index\\
            Ownership & 1: owned or being bought,  0: rented\\ 
            Gender of HH & 1: male, 2: female\\
            Race of HH & 1: white, 2: black\\
            &3: American Indian or Alaska native,\\
            & 4: Chinese, 5: Japanese,\\
            & 6: other Asian/Pacific Islander, 7: other race,\\
            & 8: two major races, 9: three or more major races\\
            &\\
            \multicolumn{2}{c}{\underline{\textit{Individual-level}}}\\
            &\\
            Within Household Index & Person indicator nested within household index.\\
            Relationship to HH & 2: spouse, 3: biological child 4: adopted  child,\\
            & 5: stepchild, 6: sibling, 7: parent 8: grandchild,\\
            & 9: parent-in-law, 10: child-in-law, 11: other relative,\\
            & 12: boarder, roommate or partner,\\
            & 13: other non-relative or foster child\\
            Gender of Individual & Same as Gender of HH variable.\\
            Race of Individual & Same as Race of HH variable.\\

            \hline
        \end{tabular}
    \label{tab:nestedVars}
\end{table}

We model the available data using the nested data  Dirichlet process mixture of multinomial distributions employed in \cite{hu2018dirichlet} as well as in \cite{akande2019simultaneous}. We model the data-generating mechanism of $\boldsymbol{X}= (X_{0}, X_{1}, X_{2}, X_{11}, X_{12}, X_{13}, X_{21}, X_{22}, X_{23})$ using
\begin{eqnarray}\nonumber
& & \hspace{-15mm} P(\boldsymbol{X} = \boldsymbol{x}|\boldsymbol{\theta})  \\ \label{eq.Nested_model}
& & \hspace{-15mm} = \sum_g \pi_g \lambda_{gx_0}^{(0)}\lambda_{gx_1}^{(1)}\lambda_{gx_2}^{(2)} 
\left(\sum_m w_{gm} 
\phi_{gmx_{11}}^{(1)} \phi_{gmx_{12}}^{(2)} \phi_{gmx_{13}}^{(3)}\right)
\left(\sum_m w_{gm} 
\phi_{gmx_{21}}^{(1)} \phi_{gmx_{22}}^{(2)} \phi_{gmx_{23}}^{(3)}\right).
\end{eqnarray}
Here, $\boldsymbol{x}=(x_0, x_{1}, x_{2}, x_{11}, x_{12}, x_{13}, x_{21}, x_{22}, x_{23})$, and $\boldsymbol{\theta}$ represents the collection of parameters defining (\ref{eq.Nested_model}), i.e., $\pi_g$, $\lambda_{gx}^{(j)}$, $w_{gm}$, and $\phi_{gmx}^{(j)}$. 

Table \ref{tab:nestedParams} displays the components of $\boldsymbol{\theta}$. These components satisfy the following constraints. We enforce  
$\sum_m w_{gm} = 1$ with $w_{gm} \ge 0$; 
for $j=0,1,2$, and each $g$, $\sum_{x \in \text{Range}(X_j)} \lambda_{gx_j}^{(j)} = 1$ and $\lambda_{gmx}^{(j)} \ge 0$;
for each $g$, $\sum_m w_{gm} = 1$ and $w_{gm} \ge 0$;
for $j=1,2$, $i=1,2$, and each $g$ and $m$, $\sum_{x \in \text{Range}(X_{ij})} \phi_{gmx}^{(j)} = 1$ and $\phi_{gmx}^{(j)} \ge 0$.

We complete the specification of the model in  (\ref{eq.Nested_model}) by assuming a prior distribution for $\boldsymbol{\theta}$ and considering a truncated version of the mixture components. Specifically, for $\pi_g$ and $w_{gm}$, we consider a prior distribution induced by a stick-breaking representation as in \cite{dunson2009nonparametric}. Since the remaining parameters (i.e., $\lambda_{gx_j}^{(j)}$ and $\phi_{gmx}^{(j)}$) lie on the simplex space, we consider Dirichlet distributions as prior distributions for these parameters. 

\begin{table}[ht]
    \centering 
    \caption{Description of parameters for nested data model}
        \begin{tabular}{l l}
            \textbf{Parameter} & \textbf{Description}\\
            \hline
            $\pi_g$ & Mixing probability for the household-level latent class\\
            $w_{gm}$ & Mixing probability for the individual-level latent class\\
            $\lambda^{(1)}_{gx_1}$& Latent class probability of ownership of house for HH $=x_1$ \\
            $\lambda^{(2)}_{gx_2}$ & Latent class probability of  HH's race $=x_2$\\ 
            $\phi^{(1)}_{gmx_{11}}$ & Latent class probability of  individual 1's gender  $=x_{11}$\\
            $\phi^{(2)}_{gmx_{12}}$ & Latent class probability of individual 1's race $=x_{12}$\\
            $\phi^{(3)}_{gmx_{13}}$ & Latent class probability of  individual 1's relationship to HH $=x_{13}$\\
            $\phi^{(1)}_{gmx_{21}}$ & Latent class probability of individual 2's gender $=x_{21}$\\
            $\phi^{(2)}_{gmx_{22}}$ & Latent class probability of individual 2's race $=x_{22}$\\
            $\phi^{(3)}_{gmx_{23}}$ & Latent class probability of individual 2's relationship to HH $=x_{23}$\\
            \hline
        \end{tabular}
        \label{tab:nestedParams}
\end{table}

\subsection{Data and Sampling Scheme}

To demonstrate the proposed latent class modeling approach under nested data, we use a sample of $N=602$ households (representing $1,806$ individuals) from the 2012 ACS obtained from \cite{akande2019simultaneous}.  We focus on five summaries of household-level and individual-level characteristics. See Table~\ref{tab:nestedSummaries} for a description and total of each count.

\begin{table}[ht]
    \centering 
    \caption{Description and counts of selected summaries.}
        \begin{tabular}{l l}
            \textbf{Description} & \textbf{Count}\\
            \hline
            $S_1$:  households with all members white & 427\\
            $S_2$: households owned by HH & 440\\
            $S_3$: households with all members of same race & 552\\
            $S_4$: households with spouse present & 401\\
            $S_5$: households with same race couple present & 376 \\
            \hline
        \end{tabular}
        \label{tab:nestedSummaries}
\end{table}

\subsubsection{Implementation  Details}

Recall from Section 3 in the main text that we treat each count as an independent draw from a binomial distribution with parameters $N$ and $P(\boldsymbol{\theta})$.  Using  (\ref{eq.Nested_model}), we can compute directly the probability for each individual combination of data and for each of the selected summaries in Table~\ref{tab:nestedSummaries}.  We have:
\begin{equation}
\begin{split}
P(S_1|\boldsymbol{\theta}) &= \sum_g \pi_g \lambda^{(2)}_{g1} \left(\sum_m w_{gm} \phi^{(2)}_{gm1} \right)^2\\
P(S_2|\boldsymbol{\theta}) &= \sum_g \pi_g \lambda^{(1)}_{g1}\\
P(S_3|\boldsymbol{\theta}) &= \sum_{r \in \mathrm{Races}} \sum_g \pi_g \lambda^{(2)}_{gr} \left(\sum_m w_{gm} \phi^{(2)}_{gmr} \right)^2\\
P(S_4|\boldsymbol{\theta}) &= 2 \sum_g \pi_g \left( \sum_m w_{gm} \phi^{(3)}_{gm2}\right) - \sum_g \pi_g \left( \sum_m w_{gm} \phi^{(3)}_{gm2}\right) ^2\\
P(S_5|\boldsymbol{\theta}) &= 2 \sum_{r \in \mathrm{Races}} \sum_g \pi_g \lambda^{(2)}_{gr} \left(\sum_m w_{gm} \phi^{(2)}_{gmr} \phi^{(3)}_{gm2} \right) - \sum_{r \in \mathrm{Races}} \sum_g \pi_g \lambda^{(2)}_{gr} \left(\sum_m w_{gm} \phi^{(2)}_{gmr} \phi^{(3)}_{gm2} \right) ^2 .\\
\end{split}
\end{equation}

Since we have access to the noisy versions of $S_1,\ldots,S_5$, we use the same model specification as in our proposed approach of the main text, where the probability of success for each count is given by the corresponding  $P(S_t|\boldsymbol{\theta})$. We use a  Metropolis-Hastings-within-Gibbs sampling algorithm was used to estimate model parameters. We set the number of household and individual latent classes (or truncated levels) equal to two. The prior distributions for the different components of $\boldsymbol{\theta}$ correspond to uniform distributions on the corresponding simplex spaces. We draw proposals for each of the probabilities from a Dirichlet distribution with concentration parameter centered on the last value of the chain. We run the algorithm for 20,000 iterations, discarding the first 10,000 iterations as burn-in. We thin the truncated chain by every five iterations. We monitor convergence of model parameters using trace plots of probabilities for each count.

We run the model under two separate schemes. First, we use the true counts without added noise to assess the performance of the estimation method without concerns about privacy. Second, we add independent noise drawn from the Geometric mechanism to each count with $\epsilon \in \{0.01, 0.1, 1.0\}$.

\subsection{Results}

\begin{figure}[htbp]
	\centering
	\includegraphics[width=4in]{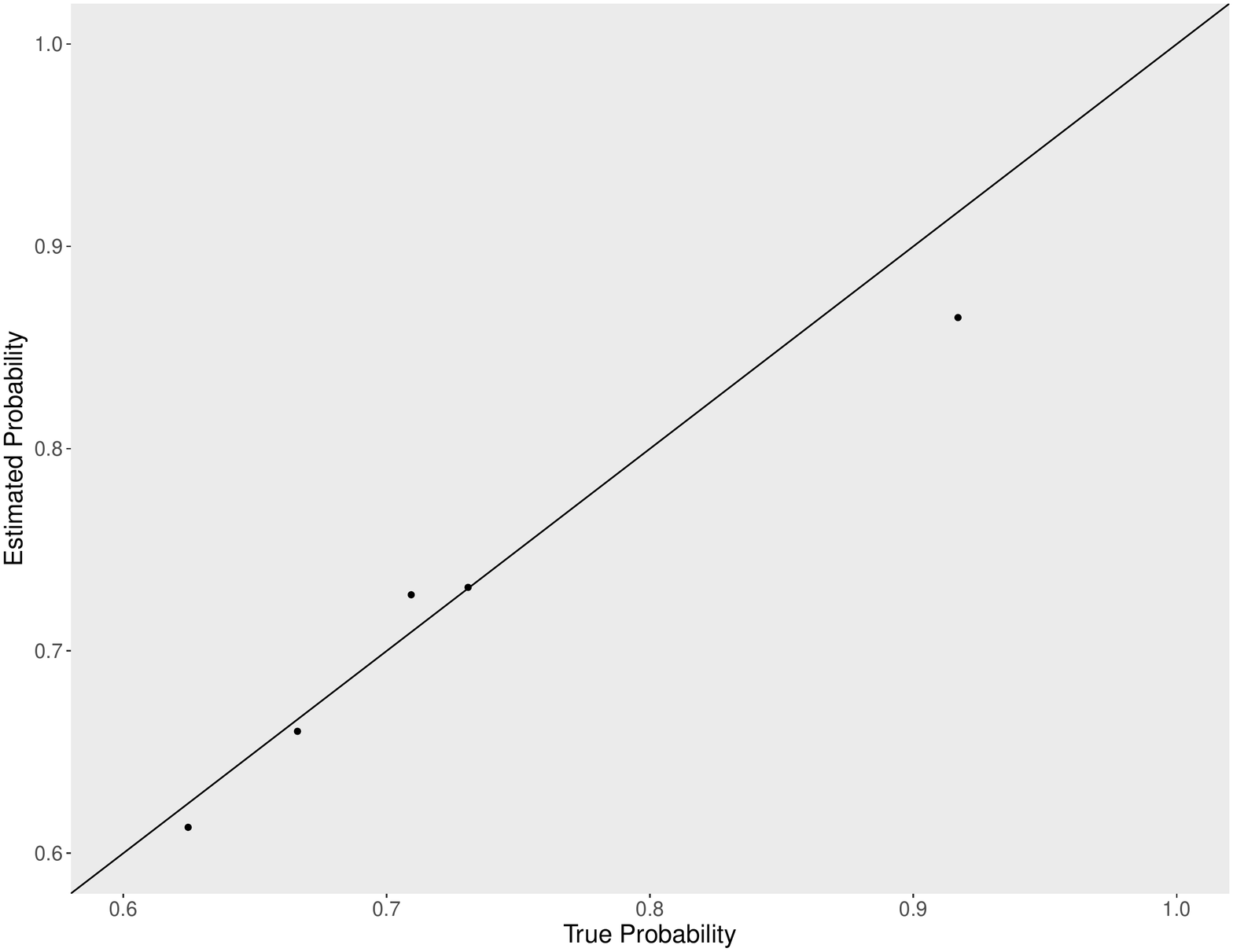}\\
	\caption{True versus estimated probabilities for the augmented nested model with no additional noise added.}
	\label{fig:nested_noNoise}
\end{figure}

We first assess the effect of the assumption of independence between counts. Figure~\ref{fig:nested_noNoise} displays the results for simulations. In general, the assumption of independence appears not to be problematic for the small number of counts used in parameter estimation. 

\begin{figure}[htbp]
	\centering
	\includegraphics[width=4in]{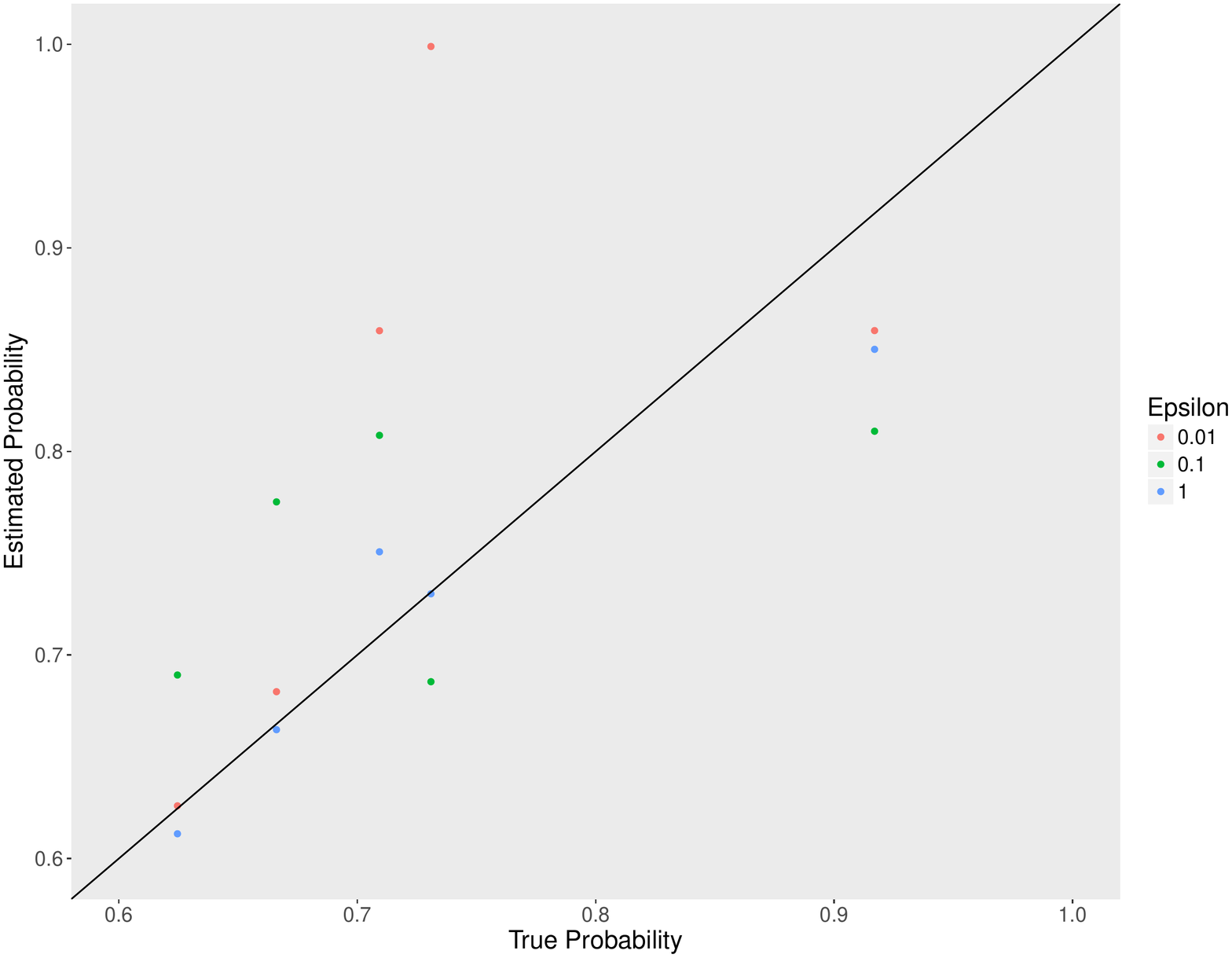}\\
	\caption{True versus estimated probabilities for the augmented nested model for a variety of values of $\epsilon$. Noise was added directly to the underlying counts to satisfy privacy.}
	\label{fig:nested_Noise}
\end{figure}

We next use the approach to generate differentially private synthetic nested data. We add privacy-preserving noise from the Geometric mechanism directly to each count. We allot one fifth of the privacy budget to each count. Results for this simulation are shown in Figure~\ref{fig:nested_Noise}. For large values of $\epsilon$, the synthetic data provide estimates that mimic those in the underlying cell probabilities. As expected, this result degrades as privacy guarantees are made stronger.

Further investigation is needed to determine the quality of inference of the synthetic data, how the choice of included counts affects the estimation of other non-included counts, and applicability to higher dimensional data sets. These are important topics for future research. 

\section{Sampling Details}

In this section, we discuss the sampling strategy. The parameters $(M_1, \ldots, M_t, \boldsymbol{\pi}_k,  \boldsymbol{\Psi}_k)$ are drawn from a posterior distribution of the form:
\begin{equation}
    f(M_1, \ldots, M_t, \boldsymbol{\pi}_k, \boldsymbol{\Psi}_k)| \Tilde{M}) \propto f(\Tilde{M} | M_1, \ldots, M_t) f(M_1, \ldots, M_t | \boldsymbol{\pi}_k, \boldsymbol{\Psi}_k) f(\boldsymbol{\pi}_k) f(\boldsymbol{\Psi}_k).\\
\end{equation}

Many Bayesian latent class models specify a Dirichlet process prior for $\boldsymbol{\pi}_k$ and Dirichlet priors for $\boldsymbol{\Psi}_k$. They rely on conjugacy to sample from the posterior distributions of $\boldsymbol{\pi}_k$ and $\boldsymbol{\Psi}_k$. Two intricacies complicate such a sampling strategy with our proposed approach. First, since the full data are not available, it is not feasible to assign latent classes for every observation. These latent class assignments are typically used in the sampling of $\boldsymbol{\pi}_k$. Second, the distribution of $f(M_1, \ldots, M_t| \boldsymbol{\pi}_k, \boldsymbol{\Psi}_k)$ is parameterized by functions of $(\boldsymbol{\pi}_k, \boldsymbol{\Psi}_k)$ and not $\boldsymbol{\Psi}_k$ directly as is typically the case.

To sample from the posterior distributions, we use a Metropolis-Hastings-within-Gibbs sampling algorithm. We iteratively update between $M_1, \ldots, M_t$, $\boldsymbol{\pi}_k$, and $\boldsymbol{\Psi}_k$ by sampling from each of the respective complete conditional distributions. At each iteration $m=1, \dots, N.runs$, we proceed as follows.

\vspace{.25in}

\noindent \textbf{S1: Sampling for $M_1, \ldots, M_t$:}

\begin{enumerate}
    \item For $t=1,...,T$:
    \begin{enumerate}
        \item  Propose $M_t^*$ from a truncated two-sided geometric distribution centered at $\Tilde{M}_t$ and scale parameter equal to $\exp \left\{ \frac{-\epsilon}{\Delta M_t T} \right\}$.
        \item Calculate the acceptance probability $\alpha= \frac{f(M_t^*)}{f(M_t^{m-1})}$ where $f(M_t|\Tilde{M},\boldsymbol{\pi}_k, \boldsymbol{\Psi}_k ) \propto f(\Tilde{M_t}|M_t) f(M_t| \boldsymbol{\pi}_k, \boldsymbol{\Psi}_k)$.
        $$
        \alpha= \frac{f(M_t^*|\Tilde{M},\boldsymbol{\pi}_k, \boldsymbol{\Psi}_k )g(M_t^{m-1}|M_t^*)}{f(M_t^{m-1}|\Tilde{M},\boldsymbol{\pi}_k, \boldsymbol{\Psi}_k )g(M_t^*|M_t^{m-1})},
        $$
        where $g()$ denotes the truncated two-sided geometric proposal distribution. 
        
        \item Generate $Y \sim \mathrm{Unif}(0,1)$. Set $M_t^{m} = M_t^*$ if $Y \le \alpha$, otherwise set $M_t^{m} = M_t^{m-1}$.
    \end{enumerate}
\end{enumerate}

\vspace{.1in}
\noindent \textbf{S2: Sampling for $\boldsymbol{\pi}_k$:}

\vspace{.1in}
\noindent
\textit{Strategy 1:}
\begin{enumerate}
    \item Propose $\boldsymbol{\pi}_k^*$ from a Dirichlet distribution parameterized by $\boldsymbol{\pi}_k^{m-1}$.
    \item Calculate the acceptance probability $\alpha= \frac{f(\boldsymbol{\pi}_k^*) g(\boldsymbol{\pi}_k^{m-1}| \boldsymbol{\pi}_k^*)}{f(\boldsymbol{\pi}_k^{m-1}) g(\boldsymbol{\pi}_k^*| \boldsymbol{\pi}_k^{m-1})}$ where $f(\boldsymbol{\pi}_k| M, \boldsymbol{\Psi}_k) \propto f(M_t| \boldsymbol{\pi}_k, \boldsymbol{\Psi}_k)) f(\boldsymbol{\pi}_k) \propto \prod_{t=1}^T {\rm Multinomial}(n,P_t(\boldsymbol{\pi}_k, \boldsymbol{\Psi}_k)) f(\boldsymbol{\pi}_k)$ and $g()$ denotes the Dirichlet proposal distribution.
    \item Generate $Y \sim \mathrm{Unif}(0,1)$. Set $\boldsymbol{\pi}_k^{m} = \boldsymbol{\pi}_k^*$ if $Y \le \alpha$, otherwise set $\boldsymbol{\pi}_k^{m} = \boldsymbol{\pi}_k^{m-1}$.
\end{enumerate}

\vspace{.1in}
\noindent
\textit{Strategy 2:}
\begin{enumerate}
    \item Reparameterize $\boldsymbol{\pi}_k$ as $\pi_i = \eta_i / \sum_i \eta_i$ for $i = 1,...,k$ for $\eta_1,...,\eta_k \in [0,\infty]$.
    \item Propose $\eta_i^*$ for $i=1.,,,k$ from a Normal distribution with mean $\eta_i^{m-1}$ and variance tuned for an appropriate acceptance rate. Reject any negative proposals by returning $\eta^{M-1}$.
    \item Calculate the acceptance probability $\alpha= \frac{f(\boldsymbol{\eta}_k^*| M, \boldsymbol{\Psi}_k)}{f(\boldsymbol{\eta}_k^{m-1}| M, \boldsymbol{\Psi}_k)}$ where $f(\boldsymbol{\eta}_k| M, \boldsymbol{\Psi}_k) \propto f(M_t| \boldsymbol{\eta}_k, \boldsymbol{\Psi}_k)) f(\boldsymbol{\eta}_k) \propto \prod_{t=1}^T {\rm Multinomial}(n,P_t(\boldsymbol{\pi}_k, \boldsymbol{\Psi}_k)) f(\boldsymbol{\eta}_k)$. Note that the finite representation of the stick-breaking prior on $\boldsymbol{\pi}_k$ implies a Gamma prior with constant scale on $\eta_i$ \citep{ishwaran2000markov,ishwaran2001gibbs}.
    \item Generate $Y \sim \mathrm{Unif}(0,1)$. Set $\boldsymbol{\pi}_k^{m} = \boldsymbol{\pi}_k^*$ if $Y \le \alpha$, otherwise set $\boldsymbol{\eta}_k^{m} = \boldsymbol{\eta}_k^{m-1}$.
\end{enumerate}

\vspace{.1in}
\noindent \textbf{S3: Sampling for  $\boldsymbol{\Psi}_k= \{\Psi_h^{(j)}\}_{h=1,j=1}^{k,p}$:}

\begin{enumerate}
    \item For $j=1,...,p$:
    \begin{enumerate}
        \item For $h=1,...,k$:
        \begin{enumerate}
            \item Propose ${\Psi}_h^{(j)*}$ from a Dirichlet distribution parameterized by ${\Psi}_h^{(j),m-1}$.
    \item Calculate the acceptance probability $\alpha= \frac{f(\Psi_k^{(j)*}) g(\Psi_k^{(j),m-1}| \Psi_k^{(j)*})}{f(\Psi_k^{(j),m-1}) g(\Psi_k^{(j)*}| \Psi_k^{(j),m-1})}$ where $f(\Psi_k^{(j)}| M, \boldsymbol{\pi}_k) \propto f(M_t| \boldsymbol{\pi}_k, \Psi_k)) f(\Psi_k^{(j)}) \propto \prod_{t=1}^T {\rm Multinomial}(n,P_t(\boldsymbol{\boldsymbol{\pi}_k}, \Psi_k)) f(\Psi_k^{(j)})$ and $g()$ denotes the Dirichlet proposal distribution.
    \item Generate $Y \sim \mathrm{Unif}(0,1)$. Set $\Psi_k^{(j),m} = \Psi_k^{(j)*}$ if $Y \le \alpha$, otherwise set $\Psi_k^{(j),m} = \Psi_k^{(j),m-1}$.
        \end{enumerate}
    \end{enumerate}
\end{enumerate}

Our algorithm returns all parameter estimates for each iteration. With these estimates, we can calculate corresponding marginal and full table probabilities. If desired, synthetic data of a desired size can be created using the full table probability estimates either through post-processing or sampling.

Our package supports both sampling strategies for $\mathbf{\pi_k}$. In practice, however, we have found that Strategy 2 offers better mixing performance.

\section{Run Time Complexity}

Most of the computational complexity in our proposed approach stems from calculating the two-way marginal probabilities, i.e., $P(\pi, \psi)$. This complexity is effected by both the dimension of the table and the number of latent classes. For computational efficiency, functions for evaluating marginal probabilities were written in base \texttt{R}. Functions were vectorized and parallelized using the \texttt{parallel} \citep{parallelPackage} and \texttt{Rfast} \citep{RfastPackage} packages where possible.

\begin{figure}[h]
    \centering
    \includegraphics[width = 3in]{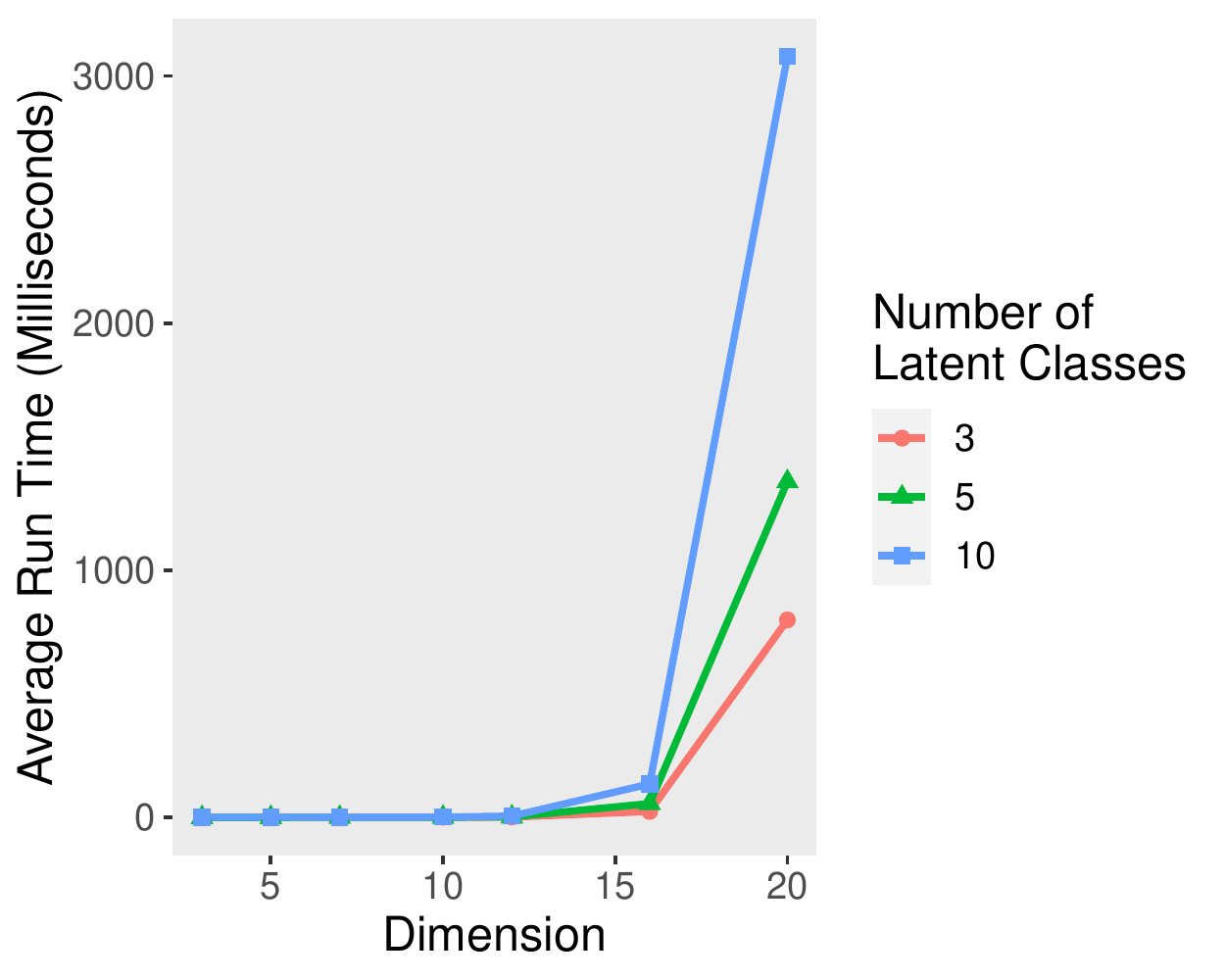}
    \caption{Average running time in milliseconds for computing a single two-way marginal probability as a function of dimension and the number of latent classes. Average computed over 100 iterations.}
    \label{fig:run-time}
\end{figure}

To show how computational cost changes as a function of dimension and the number of latent classes, we simulated a two-way table with a given size and calculated how long it took to compute a single two-way marginal probability. Code was run on a Windows laptop (Intel i7 processor with 16 GB RAM) using 7 cores. Results are shown in in Figure~\ref{fig:run-time}. Results were averaged over 100 replicates. As shown in the figure, this function performs well for moderately sized tables. However, it increases in as dimension increases, regardless of the number of latent classes.

\bibliography{references}
\bibliographystyle{abbrvnat}